\documentclass[12pt,preprint]{aastex}
\slugcomment{Accepted for publication in The Astrophysical Journal}
\shorttitle{Do we really know the dust?}
\shortauthors{Juh\'asz et al.}

\begin{document}
   \title{Do we really know the dust? Systematics and uncertainties of the mid-infrared spectral analysis methods}
   \author{A. Juh\'{a}sz\altaffilmark{1}, Th. Henning\altaffilmark{1,2},
           J. Bouwman\altaffilmark{1}, C.P. Dullemond\altaffilmark{1},
           I. Pascucci\altaffilmark{3}, D. Apai\altaffilmark{4,5}}
   \email{juhasz@mpia-hd.mpg.de}
   \altaffiltext{1}{Max-Planck-Institut f\"ur Astronomie, 
              K\"onigstuhl 17, D-69117 Heidelberg, Germany}
   \altaffiltext{2}{Kavli Institute for Theoretical Physics, Kohn Hall, University of California, 
              Santa Barbara, CA 93106, USA}
  \altaffiltext{3}{Johns Hopkins University,  3400 N. Charles St. Baltimore, MD 21218, USA}
   \altaffiltext{4}{Space Telescope Science Institute, 3700 San Martin Drive, Baltimore, MD 21218, USA}
   \altaffiltext{5}{Laplace Team of NASA Astrobiology Institute}

  \begin{abstract}
    The spectral region around 10\,{\micron}, showing prominent emission bands from various
    dust species is commonly used for the evaluation of the chemical composition of protoplanetary 
    dust. Different analysis methods have been proposed for this purpose, but so far, 
    no comparative test has been performed to test the validity of their assumptions. In this paper, we evaluate how 
    good the  various methods are in deriving the chemical composition of dust grains from infrared 
    spectroscopy.

    Synthetic spectra of disk models with different geometries and central sources were calculated, using a 2D
    radiative transfer code. These spectra were then fitted in a blind test by four spectral 
    decomposition methods.  We studied the effect of disk structure (flared vs.\ flat), inclination 
    angle, size of an inner disk hole and stellar luminosity on the fitted chemical composition.

    Our results show that the dust parameters obtained by all methods deviate systematically
    from the input data of the synthetic spectra. 
    The dust composition fitted by the new two-layer temperature distribution method, described in this paper, 
    differs the least from the input dust composition and the results show the weakest systematic effects. 
    The reason for the deviations of the results given by the previously used methods lies in their simplifying 
    assumptions. Due to the radial extent of the 10\,{\micron} emitting region there is dust
    at different temperatures contributing to the flux in the silicate feature. Therefore, the 
    assumption of a single averaged grain temperature can be a strong limitation of the previously 
    used methods. The continuum below the feature can consist of multiple components (e.g.,\ star, 
    inner rim, disk midplane), which cannot simply be described by a Planck function at a 
    single temperature. In addition, the optically thin emission of "featureless" grains 
    (e.g.,\ carbon in the considered wavelength range) produces a degeneracy in the models with 
    the optically thick emission of the disk. 
    
    The effect of different noise levels on the results has also been tested. We find that
    for a singal-to-noise ratio of 100  one can expect an \emph{absolute} uncertainty in the value of the
    crystallinity of about 11\,\% using ground based observations (8--13\,$\mu$m). 
    For space based observations (7--17\,$\mu$m) the expected uncertainty is about
    5\,\% for the same signal-to-noise value.  Moreover, the average value of the estimated crystallinity 
    increases toward lower signal-to-noise ratios in general. On the basis of our 
    results, we propose a recipe for the analysis and interpretation of dust spectroscopy data in the 
    mid-infrared which should be especially valuable for analyzing Spitzer spectroscopy data and 
    ground-based infrared spectroscopy data in the 10\,{\micron} window.

  \end{abstract}

   \keywords{astrochemistry -- infrared:general -- stars:circumstellar matter -- protoplanetary disks --
             lines:profiles}

%
\section{Introduction}

Excess emission above the stellar photosphere at mid-infrared wavelengths is a
characteristic feature of Young Stellar Objects (YSOs). This infrared excess is caused
by the thermal emission of circumstellar dust around the central star gathered
in a disk-like structure and/or envelope. The amount of excess emission and the shape of 
the Spectral Energy Distribution (SED) is indicative of the evolutionary state of the object 
(e.g.,\ \citealt{ref:van_boekel_spie}). Not only geometrical and density the structure of the 
circumstellar material, but also the dust properties evolve with time, due to, e.g., crystallization 
and coagulation (e.g.,\ \citealt{ref:beckwith,ref:henning_2006} and references therein), vaporization 
and re-condensation \citep{ref:gail_2004}.

The dust evolution and grain processing can be even more clearly seen in the 
emission/absorption features observed in the mid-infrared spectrum of many young
stars with a wide range of spectral types (e.g.,\ \citealt{ref:molster_waters, ref:apai, ref:bouwman_2008}; 
Henning 2008 (in press)). The most frequently observed features, around 10 and 18\,{\micron}, 
are caused by amorphous and crystalline silicate grains. In the case of disk emission spectra, 
they arise from optically thin regions. Therefore, their shape is directly determined 
by the optical properties of the individual dust grains. The silicate features have been 
used in order to determine the dust properties, e.g., the fraction of crystalline grains, the 
iron/magnesium ratio or the characteristic grain size in surface layers of protoplanetary disks 
(e.g.,\ \citealt{ref:waelkens,ref:bouwman_2001,ref:li,ref:van_boekel_2005a}) or 
of Solar System comets \citep{ref:crovisier_1997, ref:brucato,ref:wooden,ref:min_2005_icar}. Amorphous silicates with olivine and
pyroxene stoichiometry are common materials in the diffuse interstellar medium (ISM) as 
well as in cicumstellar disks \citep{ref:henning_2005}. The broad and smooth features 
with a peak at approximately 9.7\,{\micron} and 18.5\,{\micron} are attributed to
the Si--O stretching and O--Si--O bending modes, respectively, 
in amorphous silicate grains. Crystalline silicates (e.g., forsterite and enstatite) have 
in particular been observed in protoplanetary disks and they seem to be absent from 
the diffuse ISM. \citet{ref:kemper} placed an 2.2\,\% upper limit for the crystallinity 
fraction of the silicates in the diffuse ISM. Thus it seems to be reasonable to conclude that crystalline 
silicates form in protoplanetary disks. However, the details of the 
formation process remain unclear. Crystalline silicate grains can be produced by 
direct condensation from the gas phase \citep{ref:gail_2004} or by annealing of amorphous grains 
at higher temperatures \citep{ref:fabian_2000, ref:gail_2001, ref:harker_2002}, or both processes.

Space-based observations of disks with the {\it Infrared Space Observatory} (ISO) and the
{\it Spitzer Space Telescope} (SST) as well as ground-based measurements with, 
e.g.,\ TIMMI2, TReCS or COMICS, enabled the detailed investigation of the silicate 
features. Such studies have shown that dust processing clearly occurs in protoplanetary 
disks around stars with different 
spectral types (e.g.,\ \citealt{ref:bouwman_2001,ref:van_boekel_2005a, ref:apai, ref:kessler_silacci_06, 
ref:honda_2006, ref:sargent_2006, ref:watson_2007, ref:sicilia_aguilar_2007, ref:bouwman_2008}). 
These studies also demonstrated that evidence for grain growth can be 
found in the mid-infrared spectra, indicated by the broadening of the 10\,{\micron} emission feature
together with a decrease of its height above the underlying continuum. The relation, however, 
between grain growth and crystallization and the role of the stellar type is a topic of ongoing debate. 
There are reports on possible correlations between grain growth and crystallization 
\citep{ref:van_boekel_2005a,ref:apai,ref:kessler_silacci_06} 
and on an inverse correlation between the average grain size and stellar type 
\citep{ref:kessler_silacci_06} and a possible inverse correlation between the crystallinity and the
stellar temperature \citep{ref:apai}.

The results of these studies depend not only on the quality of observational data, but also 
on the applied analytical methods. Here, we should also note, that the features are produced
in the surface layer and the material properties are not necessary typical of the material
in the disk midplane. In addition, the region emitting the 10\,{\micron} feature will be 
located at different distances from the central star, depending on the radiation characteristics 
of the star and the nature of the dust grains and the disk properties. Due to the lack of a comparative 
study of the various analysis methods, it is practically impossible to find out if the relations, discussed above,
are really present or if they 
are caused by systematic uncertainties in the applied methods. Since we can determine the physical 
properties and chemical composition of protoplanetary dust grains for a larger sample only via such
spectral decomposition methods, it is important to know how good these methods are and which of them 
should be applied in 
a specific case. In this paper, using a 2D radiative transfer model of protoplanetary disks with a 
prescribed dust composition, we synthesize "observations" which we fit with three widely used 
spectral decomposition methods. We then compare the resulting dust composition to the original input 
composition. We also propose a new and fast method for fitting the mid-infrared spectral features in which we 
apply a distribution of temperatures, leading to more reliable results.

%

\section{Modelling}

\subsection{Disk models}
In order to investigate the quality of the spectral decomposition methods, we fitted 
synthetic spectra calculated by the 2D radiative transfer 
code RADMC \citep{ref:DD04}. Although this code assumes an axisymmetric density 
distribution, photon packages are followed in three dimensions by a Monte Carlo method. 
After the temperature distribution has been determined, the ray tracing module
of the more general code RADICAL \citep{ref:dt} has been used to compute the SEDs. 
These codes have already been used to model protoplanetary disk SEDs in numerous studies 
\citep{ref:pontoppidan_2005, ref:van_boekel_2005b,ref:DD04} and
tested against other continuum radiative transfer codes \citep{ref:pascucci}.

The disk structure can possibly be related to grain evolution. A flared disk, containing
unprocessed, small amorphous grains becomes flattened with time as dust grains
grow and settle to the disk midplane \citep{ref:bouwman_2008}. Since the flaring angle of the disk has a strong
effect on the shape of the continuum at mid-infrared wavelengths \citep{ref:DD04}, it can
also have an effect on the results of the spectral analysis methods. Therefore,
we used disk models with different geometries. We want to note that we did not
treat the dust coagulation and sedimentation in a self-consistent way during the
modelling. Since the irradiated area of a 
circumstellar disk depends on the luminosity of the central star, we used stars with 
three different types (Herbig Ae star, T Tauri star and Brown Dwarf) as central 
radiation source in our models. The main parameters of the 2D RT disk models as well
as the applied stellar properties are summarized in Tab.\,\ref{tab:mod_prop}. 
The stellar emission has been modelled by a blackbody radiation source instead of a Kurucz 
model \citep{ref:kurucz}, which made the models more uniform enabling us to compare the 
results directly.

We calculated six series of disk models in order to investigate the effect of disk 
structure and disk orientation on the results of the fitting procedure. Each of the model 
series contained 12 spectra: three different stellar types and four different inclinations 
from pole-on to close to edge-on orientation (0$^{\circ}$, 20$^{\circ}$, 45$^{\circ}$, 70$^{\circ}$). 
The mass of the disk in each model has been scaled with the stellar mass as $M_{\rm{disk}}=0.03M_{\star}$.
The surface density as a function of radius was described by a power law as 
$\Sigma (r)=\Sigma_0(r/r_{\rm disk})^p$, where $\Sigma_0$ is the surface density
at the outer disk radius and $p=-1$. 
The outer radius of the Herbig Ae disk was set to 400\,AU. Then the outer disk 
radius of the T Tauri and Brown Dwarf disks was determined by the radial optical 
depth at 0.55\,{\micron}, which we took to be same as in the Herbig Ae disk. It is important
to note that the chemical composition of dust grains was assumed to be uniform throughout the disk. 
This simplifying assumption will be less justified when considering a large wavelength range
and, therefore, a wider temperature and radial region in the disk. 
The passive irradiated disk model of \citet{ref:ddn2001} has been applied 
for the disk structure in general, but the detailed structure of the disks were
different in the different model series. The six model series can be summarized 
as follows.

{\bf Model AS, AB.} A flared disk geometry has been applied with a flaring 
index $\alpha$ (see Eq.\,\ref{eq:hr}) of 2/7 and with a dimensionless pressure
scale height at the inner radius of $H(R_{in})/R_{in}=0.03$ 
(see Appendix\,\ref{sec:app_disk_models}). The inner disk radius was 
determined by the sublimation temperature of 1500\,K (Model AS) and 
1060\,K (Model AB). The models with 1060\,K temperature at their inner edge have
twice as large inner holes as the models with 1500\,K temperature.

{\bf Model BS, BB.} Here we applied a moderately flared disk with a flare 
index of 1/7 and with a dimensionless pressure scale height at the inner
radius of $H(R_{in})/R_{in}=0.03$. The inner disk radius is again determined 
by the grain temperature of 1500\,K (Model BS) and 1060\,K (Model BB). 

{\bf Model CS, CB.} A more flattened disk model has been considered here compared to
the Model B series. We used a similar density structure for these models as it 
was used for Model B disks. For Model C the flaring index was set to 1/8 
producing a self-shadowed flattened disk structure. The inner disk radius was again set 
by the grain temperature of 1500\,K (Model CS) and 1060\,K (Model CB).

\subsection{Mass absorption coefficients}

During the tests, we used six different dust species (amorphous silicates with olivine and
pyroxene stoichiometry, forsterite, enstatite, silica and carbon), each with three grain 
sizes (0.1, 1.5 and 6.0\,{\micron}). Tab.\,\ref{tab:species} summarizes the used dust species, 
grain model and origin of the optical constants. The relative mass fraction for each
dust species and grain size used in the study is presented in Tab.\,\ref{tab:results}.
The used relative mass fractions as a function of grain size corresponds to number density 
distribution of  $n\propto a^{-3.5}$ and $n\propto a^{-3.3}$ for the amorphous and crystalline
dust species, respectively. 
The mass absorption coefficients of these dust species as a function of wavelength 
are shown in Fig.\,\ref{fig:opacities}. In order to calculate the mass absorption
coefficient from the optical constants, we used the theory of Distribution of Hollow 
Spheres (DHS) \citep{ref:min_2005} for crystalline silicates, while classical Mie theory 
for spherical particles was used for the amorphous dust species. 

There are two distinct, but coupled problems related to the fitting of the 
mid-infrared silicate features. One of the problems is the applicability of a special set of
optical data and the other one is the uncertainty in the applied spectral 
decomposition methods, themselves. It is unknown whether or not we use exactly the right 
optical data, which determine how the protoplanetary dust grains interact with radiation.
The question of how good our set of mass absorption coefficients is is beyond 
the scope of this paper. Here, we used the same dust composition 
for all disk models (see Tab.\,\ref{tab:results}), in order to evaluate the performance of the
various spectral decomposition methods. We are aware that the use of one specific
dust composition in all disk models is a limitation of the study, but it is impractical  
to test the methods on all possible combinations of dust mixtures and stellar/disk parameters. 
We chose, therefore, a single dust composition for the disk models the calculated spectra of which 
resembles that of typically seen towards young stellar objects. 
In this study we investigate how the different stellar and disk parameters can modify the results of 
spectral decomposition. The best way to isolate these effects is to use one 
dust mixture in all disk models, thus any difference between the input and the resulting
dust parameters is certainly caused by the effects of stellar or disk parameters,
which is then a deficiency of the applied method.

\subsection{Fitting methods}

Numerous techniques have been applied for the spectral decomposition of mid-infrared 
silicate features (e.g.,\ \citealt{ref:bouwman_2001, ref:li, ref:van_boekel_2005a,
ref:kessler_silacci_06, ref:watson_2007, ref:bouwman_2008}), 
but no systematic comparison of various techniques has been performed so far, leaving uncertainties
in the quality and applicability of these techniques. 
We choose the three most widely used methods and tested them 
against a new method described in this paper and applied them to the synthetic spectra of the 
above described disk models. These three methods have been applied by many authors 
to evaluate the chemical composition of protoplanetary dust from mid-infrared 
silicate features for a wide range of source types (e.g.,\ \citealt{ref:bouwman_2001, ref:meeus_2003,
ref:van_boekel_2005a, ref:apai, ref:honda_2006, ref:schegerer_2006, ref:kessler_silacci_06, 
ref:sicilia_aguilar_2007, ref:bouwman_2008}).
All methods assume that the silicate emission comes from an 
optically thin region, but the temperature structure and underlying continuum is 
described in different ways in the distinct methods. 

The simplest method for the spectral decomposition of the silicate emission feature is a 
continuum subtraction (hereafter COS) method (e.g.,\ \citealt{ref:bouwman_2001}).
In this approach the continuum is fitted by a polynomial (usually first or second order) 
outside the feature and then subtracted from the spectrum. In the 5--36\,{\micron}
domain ({\it Spitzer} IRS) there are three possible wavelength ranges in which 
the continuum can be fitted; shortward of 7.5\,{\micron}, between 12 and 15\,{\micron} 
and longward of 30\,{\micron}. The wavelength domains for the continuum fitting are not 
strictly determined. In this study we chose the following domains for the
continuum fitting 6.8--7.5\,{\micron}, 12.5--13.5{\,{\micron}} and 30--36\,{\micron}.
A reasonable way to subtract the continuum has been developed by 
\citet{ref:van_boekel_2005a}. Here the subtracted \emph{and} normalized flux 
is given by 
\begin{equation}
F_\nu^{{\rm norm}}= 1 + \frac{F_\nu-F_\nu^{{\rm cont}}}{<F_\nu^{{\rm cont}}>}.
\label{eq:cs_norm}
\end{equation} 
Here, $F_\nu$ is the flux density of the measured spectrum, $F_\nu^{{\rm cont}}$ is the 
flux density of the fitted continuum and $<F_\nu^{{\rm cont}}>$ is the average value of 
the continuum flux density over all the frequencies in the fitted domain. For 
the spectral decomposition problem one can write
\begin{equation}
F_\nu^{{\rm norm}} = \sum_{i=1}^N\sum_{j=1}^MC_{i,j}\kappa_{i,j}^{{\rm norm}},
\label{eq:cs}
\end{equation}
where $N$ is the number of dust species and $M$ is the number of different grain sizes, 
$\kappa_{i,j}^{\rm norm}$ is the normalized mass absorption coefficient of the dust species 
$i$ and grain size $j$. The mass fraction of a specific dust species can be calculated from
the $C_{i,j}$ values as 
\begin{equation}
m_{i,j} = \frac{C_{i,j}}{\sum_{i=1}^N\sum_{j=1}^MC_{i,j}}.
\label{eq:mfrac_def}
\end{equation}
The mass absorption coefficients should be normalized exactly the same way as it has been 
performed for the observed spectrum. In this fitting approach $\kappa_{i,j}$ is given, while 
$C_{i,j}$ has to be fitted. 

A further step was taken by adopting a single black body function as the continuum.
In this single-temperature (hereafter ST) approach (e.g.,\ \citealt{ref:van_boekel_2005a}) the continuum 
and spectral features are fitted simultaneously and the observed flux density at a specific 
frequency is given by 
\begin{equation}
F_\nu = C_0B_\nu(T) + \sum_{i=1}^N\sum_{j=1}^MC_{i,j}B_\nu(T)\kappa_{i,j}.
\label{eq:st}
\end{equation}
Here $\kappa_{i,j}$ is the mass absorption coefficient of the dust species $j$ 
and the grain size $i$, $T$ is the temperature in the continuum \emph{and} in the 
feature. $B_\nu(T)$ is the Planck function at a temperature $T$, the quantities
$N$ and $M$ denote the total number of different dust species and grain sizes, 
respectively. The mass fraction of a specific dust species is given by Eq.\,\ref{eq:mfrac_def}.
In this spectral decomposition problem $\kappa_{i,j}$ is given, while 
$T$, $C_0$ and $C_{i,j}$ have to be fitted. An obvious limitation of this method is
its simplistic treatment of the temperature structure of the radiating dust. 

The subsequent evolution in data quality allowed the adoption of more realistic models to fit
the 10\,{\micron} silicate feature. By fitting the continuum and the feature temperature in
Eq.\,\ref{eq:st}. separately one can obtain a more proper analyzing method.
In this two-temperature (hereafter TT) approach \citep{ref:bouwman_2008} the flux density is 
given by
\begin{equation}
F_\nu = C_0B_\nu(T_c) + \sum_{i=1}^N\sum_{j=1}^MC_{i,j}B_\nu(T_f)\kappa_{i,j}.
\label{eq:tt}
\end{equation}
Here $T_c$ is the temperature in the continuum and $T_f$ is the temperature 
in the feature. In all the COS, ST and TT methods,
we used a generalized version of the non-negative least-square optimization routine of 
\citet{ref:lawson_hanson}. For the ST and TT methods a grid of temperatures was created first. 
Then the optimization routine evaluated the best fitting chemical composition for each
temperature. During the tests the temperature step size was 10\,K in the 
feature and 15\,K in the continuum. The minimum and maximum temperatures in
the continuum was 20\,K and 4000\,K, respectively. For the feature we used 20\,K 
and 1500\,K as the lowest and highest temperature, respectively. We assumed that dust grains
evaporate at higher temperatures, while the lowest temperature was set to the temperature
of the interstellar dust. In order to characterize the quality of the
fit, we calculated the $\chi^2$ value for each temperature, with
\begin{equation}
\chi^2=\frac{1}{N_{\nu}-f}\sum_{i=1}^{N_{\nu}}\frac{\left(F_{\nu}^{\rm{model}}-
F_{\nu}^{\rm{observed}}\right)^2}
{(F_{\nu}^{\rm{error}})^2}.
\label{eq:chisq_def}
\end{equation}
Here $N_{\nu}$ is the number of frequencies or wavelengths and $f$ is the number of
fitted parameters. $F_{\nu}^{\rm{observed}}$ is the flux density calculated by the 2D RT code, 
$F_{\nu}^{\rm{model}}$ is the fitted flux density by the spectral decomposition method, 
and $F_{\nu}^{\rm{error}}$ is the uncertainty in $F_{\nu}^{\rm{observed}}$ and not the 
uncertainty of the 2D RT calculation.
We simulated observations by calculating synthetic spectra and we used simulated
observational flux uncertainties by $F_{\nu}^{\rm{error}}= 0.001\cdot F_{\nu}^{\rm{observed}}$. 
Then the best fit chemical composition was the model, which had the 
lowest $\chi^2$ value. Except for Sec.\,\ref{sec:noise_level} we did not add noise
to the spectra, therefore the fitted dust parameters do not depend on the 
value $F_{\nu}^{\rm{error}}$. 

The basic limitation of all the above described methods is the simple handling of the
continuum and temperature of the radiating dust. Studies of protoplanetary disks 
have shown, that the silicate emission features arise from the uppermost optically thin disk 
layer \citep{ref:cg97, ref:menshchikov_1997} and the silicate emission zone can extend from less 
than 1\,AU to a few tens of AUs in Herbig Ae systems \citep{ref:van_boekel_2005b}. Since the 
temperature in such wide range of radii should not be constant, it seems to be reasonable to apply 
a distribution of temperatures to model the 10\,{\micron} silicate feature and the underlying 
continuum. 

Here, we propose such a new method, which is more 
sophisticated compared to the above described models, but simple enough to be applied for 
large data sets. In this method we use a simple two-layer temperature distribution (hereafter TLTD 
method) in order to describe the temperature structure of the disk. In this TLTD
approach the flux density at a specific frequency is given by
\begin{equation}
F_\nu = F_{\nu,\rm{cont}} + \sum_{i=1}^N\sum_{j=1}^MC_{i,j}\kappa_{i,j}
\int_{\rm{Rrim}}^{\rm{Rout}}\frac{2\pi r}{d^2}B_\nu(T_{\rm{atm}}(r))dr, 
\label{eq:fit_td}
\end{equation}
where
\begin{eqnarray}
F_{\nu,\rm{cont}} &=& C_0\frac{\pi R_{\star}^2}{d^2}B_\nu(T_{\star}) + 
C_1\int_{\rm{Rin}}^{\rm{Rrim}}\frac{2\pi r}{d^2}B_\nu(T_{\rm{rim}}(r))dr + \nonumber \\ 
 & & C_2\int_{\rm{Rrim}}^{\rm{Rout}}\frac{2\pi r}{d^2}B_\nu(T_{\rm{mid}}(r))dr.
\label{eq:fit_td_cont}
\end{eqnarray}
Here $R_{in}$ is the inner radius of the disk, $R_{rim}$ is the outer radius of 
the puffed-up inner rim, $R_{out}$ is the outer radius of the disk, while $r$ 
is the distance to the central star and $R_{\star}$ is the radius of the star. 
$T_{\star}$ is the effective temperature of the central star, while $T_{rim}(r)$, 
$T_{atm}(r)$ and $T_{mid}(r)$ are the temperature of the inner rim, disk atmosphere
and the disk midplane, respectively. The mass fraction of a specific dust species 
is given by Eq.\,\ref{eq:mfrac_def}. The temperatures in the disk are given by
\begin{eqnarray}
T_{\rm{rim}}(r) &=& T_{\rm{rim,max}}\left(\frac{r}{R_{\rm{in}}}\right)^{q_{\rm{rim}}}\\
T_{\rm{atm}}(r) &=& T_{\rm{atm,max}}\left(\frac{r}{R_{\rm Rim}}\right)^{q_{\rm{atm}}}\\
T_{\rm{mid}}(r) &=& T_{\rm{mid,max}}\left(\frac{r}{R_{\rm Rim}}\right)^{q_{\rm{mid}}}
\end{eqnarray}
In this fitting approach $R_{\star}$, $T_{\star}$ are given, while $T_{\rm{rim,max}}$, 
$T_{\rm{mid,max}}$, $T_{\rm{atm,max}}$, $q_{\rm{rim}}$, $q_{\rm{mid}}$,
$q_{\rm{atm}}$,$C_0$, $C_1$, $C_2$ and $C_{i,j}$ have to be fitted. 
Apart from these parameters we also need to account for three radii 
 ($R_{\rm{in}}$, $R_{\rm{rim}}$, and $R_{\rm{out}}$). 

Although there are three radii there are only two free parameters, namely their
ratios $R_{\rm{out}}/R_{\rm{rim}}$ and $R_{\rm{rim}}/R_{\rm{in}}$.  The reason
for that is the following. The replacement of boundaries in Eq.\,\ref{eq:fit_td} and Eq.\,\ref{eq:fit_td_cont} 
from $R_{\rm in}$ to $\alpha\cdot R_{\rm in}$ and from $R_{\rm out}$ to $\alpha\cdot R_{\rm out}$
will not change the resulting dust composition. This behaviour is due to the 
$C_i$ and $C_{i,j}$ factors, which become $\alpha^2\cdot C_i$ and $\alpha^2\cdot C_{i,j}$ in
the above described case. Since the mass fractions of different dust species is 
defined by Eq.\,\ref{eq:mfrac_def}, $m_{i,j}$ does not depend on $\alpha$, although 
the total dust mass does. 
This means that, in a certain sense,  the actual values of $R_{\rm in}$ and $R_{\rm out}$ are irrelevant, 
and the important parameter is rather the ratio between them. Since 
we do not apply radiative transfer calculations in the TLTD method, the radius and the temperature are
not coupled to each other in a self consistent way and the role of the boundaries in Eq.\,\ref{eq:fit_td} and 
Eq.\,\ref{eq:fit_td_cont} is only the weighting of different temperatures.

Thus, one can constrain $R_{\rm{out}}/R_{\rm{rim}}$ and $R_{\rm{rim}}/R_{\rm{in}}$  using the
reasonable assumption that the outer radius of the disk is larger than the outer radius of the silicate 
emitting region. In this case, for each value of $q_{\rm atm}$ one can calculate the $
R_{\rm{out}}/R_{\rm{rim}}$ where the contribution of the outermost annulus to the total flux at 
10\,{\micron} equals to 0.1\,\% assuming a constant value for $R_{\rm{rim}}$ (the same holds for 
$q_{\rm rim}$ and $R_{\rm{rim}}/R_{\rm{in}}$). One can then neglect the flux contribution of the annuli 
behind this radius.

This can be the most easily done by calculating the integrals in Eq.\,\ref{eq:fit_td}--\ref{eq:fit_td_cont} 
over temperatures instead of radii.
Thus, we can rewrite Eq.\,\ref{eq:fit_td} in the form of
\begin{equation}
F_\nu = F_{\nu,\rm{cont}} + \sum_{i=1}^N\sum_{j=1}^MD_{i,j}\kappa_{i,j}
\int_{\rm{T_{atm,max}}}^{\rm{T_{atm, min}}}\frac{2\pi}{d^2}B_\nu(T)T^{\frac{2-{\rm qatm}}{{\rm qatm}}}dT, 
\label{eq:fit_td_temp}
\end{equation}
where
\begin{equation}
D_{i,j} = \frac{q_{\rm atm}}{{\rm T_{atm,max}}^{-2/q}{\rm R_{\rm in}}^2}C_{i,j}
\end{equation}
The value of $R_{\rm in}$ is not important here, its role is only a dimensional 
matching, therefore, we can choose it arbitrarily, for instance to 1\,AU. 
Eq.\,\ref{eq:fit_td_cont} can also be written in the same manner. 
The upper limits of the temperatures  ($T_{\rm rim,max}$,
$T_{\rm atm,max}$ and $T_{\rm mid,max}$) are fitted. Using the above described assumption, 
one can calculate the minimum temperature for each component ($T_{\rm rim,min}$,
$T_{\rm atm,min}$ and  $T_{\rm mid,min}$), where the contribution of the corresponding annulus to the total
flux at 10\,{\micron} equals to 0.1\%. The mass fraction of a given dust species is given by 
$m_{i,j}=D_{i,j}/\sum_{i=1}^N\sum_{j=1}^MD_{i,j}$.
In order to compare the temperature distribution estimated by the TLTD method to
the calculated distribution with the 2D RT code, we used Eq.\,\ref{eq:fit_td} and 
Eq.\,\ref{eq:fit_td_cont} during the tests and the same $R_{\rm{in}}$ and $R_{\rm{out}}$ 
values were used in the fitting routine as in the 2D RT code.

The fitting of the temperatures at the boundaries ($T_i$) and the power law indices 
($q_i$) is a highly non-linear problem which is not straightforward to solve. Beside 
the non-linear nature of the problem, the $\chi^2$ surface can have numerous local 
minima, which makes the fit even more difficult. Most of the fitting routines 
like, e.g., {\tt amoeba} \citep{ref:press} are not able to find the global minimum 
or/and are very sensitive to the initial values. Therefore, we used the genetic 
optimization algorithm {\tt pikaia} \citep{ref:charbonneau} in order to fit the 
temperatures and the power indices. It has been shown that this algorithm can 
efficiently find the global minimum even in a high dimensional parameter space with 
multiple local minima \citep{ref:charbonneau}. This algorithm performs the 
maximization of a user defined function, thus we used the $g(T,q)=1/\chi^2$ 
function for that purpose. Once the temperature distribution 
is fixed, the evaluation of the $C_{i,j}$ values is a linear problem. Thus,
we used for that purpose the generalized linear least-square fitting routine 
of \citet{ref:lawson_hanson}. 

\section{Results}
%
%

In the following, we investigate the quality of the decomposition methods. We fitted
our 2D RT model spectra between 7 and 14\,{\micron} since this is a widely used
wavelength range for spectral decomposition in the case of {\it Spitzer} observations.
In order to measure the quality of the fits we calculated the mass-averaged grain size 
of the amorphous material and the crystallinity fraction from the chemical composition 
estimated by the fitting methods. We use these parameters, since grain growth and crystallisation
are among the key processes occurring during the evolution of protoplanetary dust. 
The evolutionary status of dust grains is, therefore, usually characterized by these
two parameters which are commonly derived from the observations.
For a more straightforward analysis of the results, we defined a normalized crystallinity 
as the fraction of silicate mass in crystals. The total crystallinity is the fraction of 
total dust mass in crystalline silicates. The advantage of the normalized crystallinity is 
that, for a given, fixed dust composition, its value does not change if we exclude carbon 
from the fitting procedure. 

\subsection{Disk structure and stellar type}

In the following, we investigate the effect of the disk structure and the stellar type on the 
fitted dust composition. There are three questions we address in this section. (i) How large
is the \emph{average} deviation of the dust parameters fitted by a specific method from 
the input parameters? (ii) How large is the scatter around the mean dust parameters? 
(iii) Is there any correlation, within the scatter, between the dust composition and the flaring 
index, the inclination angle or the stellar type? In order to quantify the answers for these
questions we compute the mean value, the standard deviation around the mean and the Pearson's
correlation coefficient between the chosen parameters and flaring index, inclination angle
and the stellar temperature. 
The Pearson's correlation coefficient ($r$), which is frequently used
to measrue the linear dependence of two variables ($x,y$), is calculated as
\begin{equation}
r = \frac{n\sum x_iy_i - \sum x_i\sum_{i}y_i}
{\sqrt{n\sum x_i^2 - (\sum x_i)^2}\sqrt{n\sum y_i^2 - (\sum y_i)^2}},
\label{eq:pearsonsr}
\end{equation}
where $n$ is the number of data points. 
In order to investigate the significance of the Pearson's correlation coefficient, one can
calculate the probability that the observed relation can be produced by random distribution
with the same sample size ($n$)
\begin{equation}
p(r,n) = \frac{2\Gamma\left(\frac{n-1}{2}\right)}{\sqrt{\pi}\Gamma\left(\frac{n-2}{2}\right)}
\int_{|r|}^1(1-u^2)^{(n-4)/2}du
\label{eq:ttest_r}
\end{equation}
(see, e.g., \citealt{ref:taylor})

\begin{figure*}
\includegraphics[width=8.2cm]{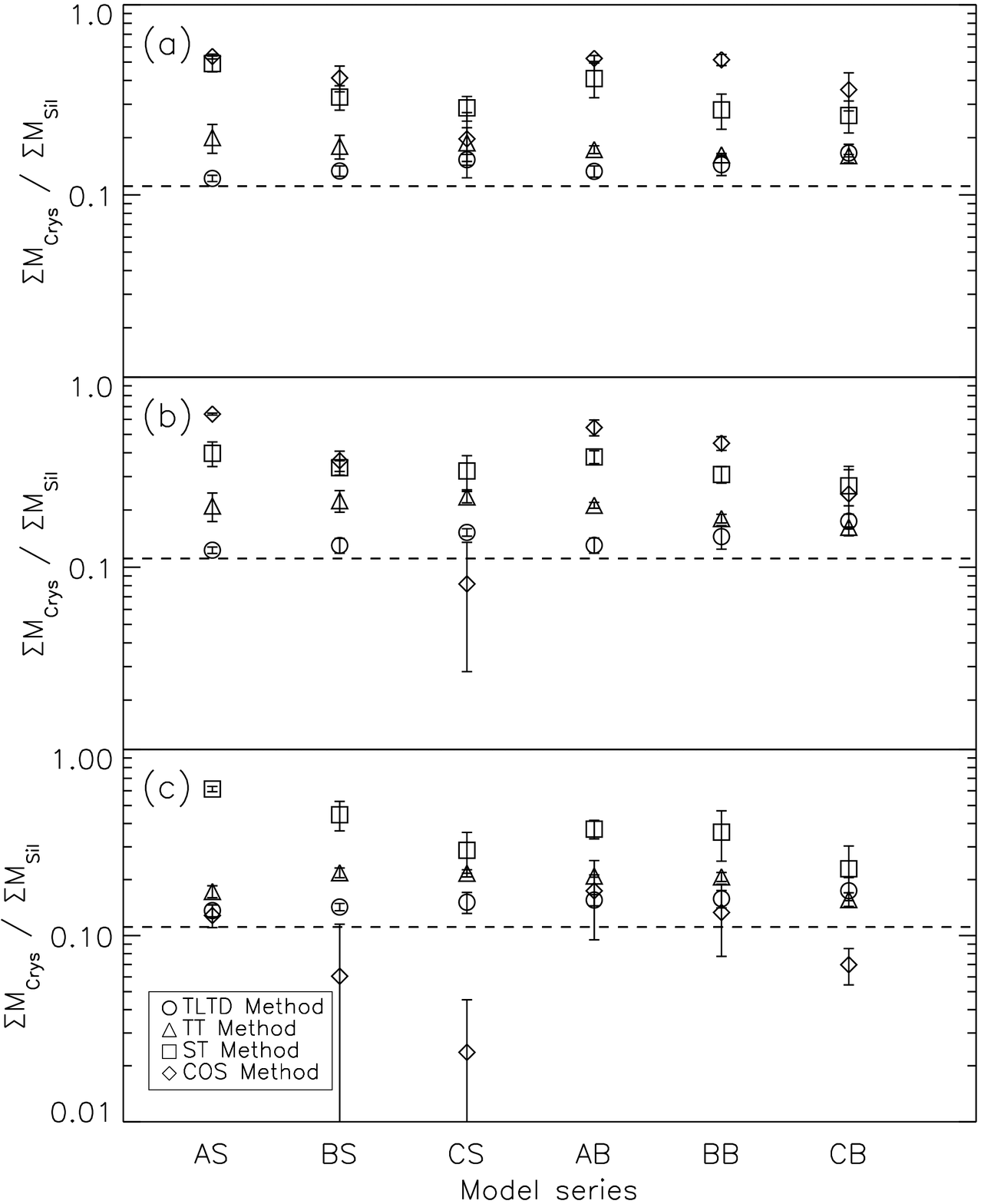}
\includegraphics[width=8.2cm]{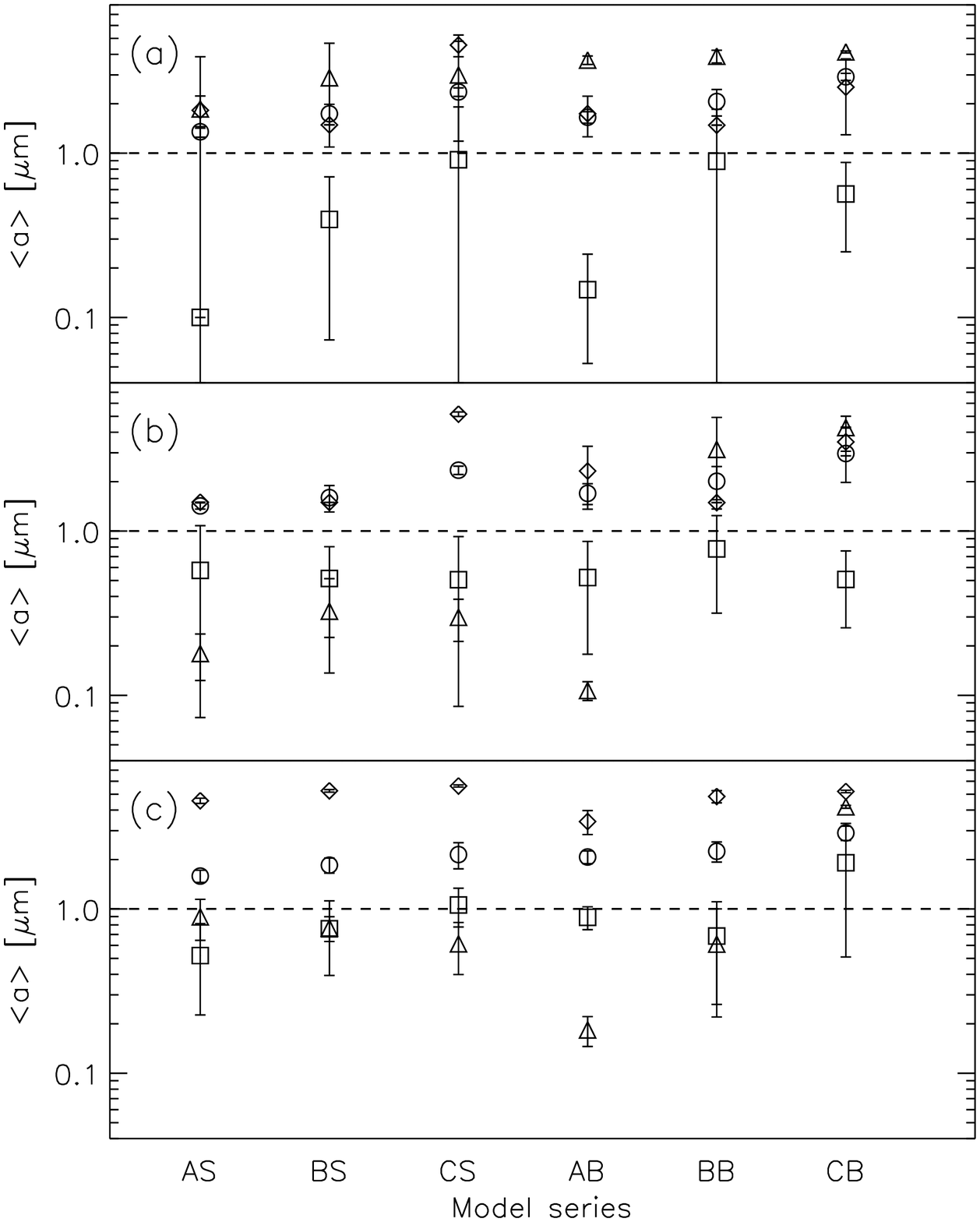}
\caption{Fraction of silicate mass in crystals ({\it Left}) and  mass-averaged
grain size ({\it Right}) for ({\it a}) Herbig Ae star, ({\it b}) T Tauri star and 
({\it c}) Brown Dwarf with carbon among the fitted dust species.
The symbols show the averaged value over the different inclination angles, and the
error bars show the standard deviation. The dashed lines show the input value to the 
RT code.}
\label{fig:comp_carbon_in}
\end{figure*}

The normalized crystallinity values obtained by the COS method show a large scatter, overestimating
and (in a few cases, e.g.,\ for Brown Dwarf in CS series) underestimating the real value 
(see Fig.\,\ref{fig:comp_carbon_in}\,{\it Left}). 
On average, the normalized crystallinity is overestimated by a factor of 2.8$\pm$1.9 and it 
correlates with the stellar temperature (r=0.53, p=$10^{-6}$) and  with the 
flaring index (r=0.47, p=$5\cdot 10^{-5}$). 
The ST method overestimates the normalized crystallinity (by a factor of 4$\pm$1.1), which
does not depend on the stellar temperature, but it correlates directly with the flaring index 
(r=0.63,  p=$3\cdot 10^{-9}$). There is also a hint of a weak inverse correlation between the crystallinity and the 
inclination angle (r=-0.3, p=0.01).
The TT method overestimates the normalized crystallinity by a factor of 1.8$\pm$0.25, which depends
weakly on the stellar temperature (r=-0.34, p=0.003) but is independent of the flaring index or of
the inclination angle. 

The total and normalized crystallinity fraction estimated by the TLTD 
method is the closest to the input value in general. This method overestimates the crystallinity 
by a factor of 1.2$\pm$0.13. The crystallinity does not depend on the stellar type but it depends 
weakly on the flaring index (r=-0.47, p=$2\cdot 10^{-5}$). The effect of the inclination on the fitted dust 
composition is the weakest for the TLTD method. Within the studied range of inclination 
angles ($0^\circ$--$70^\circ$) the scattering in the crystallinity due to the inclination angle
is less than 1\,\% for the TLTD method. It can also be seen in Fig.\,\ref{fig:comp_carbon_in}\,{\it Left} 
that the inclination has the smallest effect on the results of the TLTD method, indicated by the 
error bars, compared to other methods.

Although, the input dust composition to the 2D RT disk models did not contain large 
(6\,{\micron}) enstatite grains at all, the bulk of the crystalline silicates, 
predicted by the TT method ($\sim$60\,\% in terms of the total crystal mass), is in the form of large enstatite grains. 
The TLTD method also suffers from the overestimation of the large enstatite fraction, 
while the ST method has this problem just in a few cases. The overestimation of large 
enstatite content is probably not related to the type of the decomposition method, 
since all the methods have this problem, but it is probably related to the similarity
between the optical data of large enstatite and other dust species. The mass absorption 
coefficients of the large enstatite grains can be reproduced, in the fitted wavelength range, 
by a linear combination of small olivine, small carbon and medium sized enstatite grains. 
If this problem is really caused by the degeneracy of the optical data, we should get better 
results if we use a broader wavelength interval for the fits. The probability, that two dust 
species have similar optical data, is lower for a broader wavelength range. We will discuss 
the effect of the fitting range in more detail in Sec.\,\ref{sec:fit_range}.

The mass-averaged grain size of the amorphous silicates has also been calculated from
the fit results (Fig.\,\ref{fig:comp_carbon_in}\,{\it Left}). Since this quantity can change
by more than an order of magnitude (0.1--6.0\,{\micron}) we used the standard deviation of the
logarithm of grain size instead of the grain size itself to quantify the average and scatter of this
parameter. The COS method overestimates this quantity in all cases by a factor of 2.6$^{+2.0}_{-1.1}$.
This fitting approach predicts larger grains for lower stellar temperatures (r=-0.53, p=$5\cdot 10^{-5}$) and
for flatter disks (r=-0.46, p=$5\cdot 10^{-5}$), but the results do not depend on the studied range of 
inclination angles ($0^\circ<i<70^\circ$). 
The ST method underestimates the mass-averaged grain size by a factor of 0.42$^{+0.7}_{-0.3}$. 
The results of the ST method depend only on the stellar temperature, predicting larger
grains for lower stellar temperatures (r=-0.51, p=$4\cdot 10^{-6}$). The TT method also underestimates 
the average grain size for T Tauri and Brown Dwarf disk models, while the fitted grain size is larger than 
the input value for Herbig Ae spectra. On average the TT method underestimates the average
grain size by a factor of 0.7 $^{+1.9}_{-0.5}$, showing a larger scatter around the mean value and
correlating weakly with the stellar temperature (r=0.51, p=$4\cdot 10^{-5}$). The TLTD method overestimates the
mass-average grain size of the amorphous material by a factor of 1.8 $^{+0.4}_{-0.4}$ and the
grain size also correlates with the flaring index (r=-0.6, p=$10^{-8}$). It is interesting to note, that this
correlation decreases towards lower stellar temperatures (r=-0.7, r=-0.6 and r=-0.3 for
Herbig Ae, T Tauri and Brown Dwarf models, respectively).

\subsection{Featureless grains}
The amount of "featureless" grains (like, e.g., carbon) can also be important in evaluating
the fraction of crystalline grains and mass-averaged grain size. Due to their smooth, featureless
opacity curve in the fitted region (7--14\,{\micron}), the optically thin emission of such 
grains acts like a continuum. Thus, there is a degeneracy between the optically thin emission of 
featureless grains and the optically thick emission of different parts of the disk. This degeneracy 
can have a strong effect on the results of the fits. The importance of this degeneracy is well
reflected by the fact, that the ST method estimated a very high fraction of amorphous carbon (up to 90\,\%
in terms of total dust mass) in all test cases. The TT method also had this problem in several cases, while the TLTD method 
did not show such a behavior. We discuss this behaviour in more detail in 
Sec.\,\ref{sec:discussion}.

\begin{figure*}
\includegraphics[width=8.2cm]{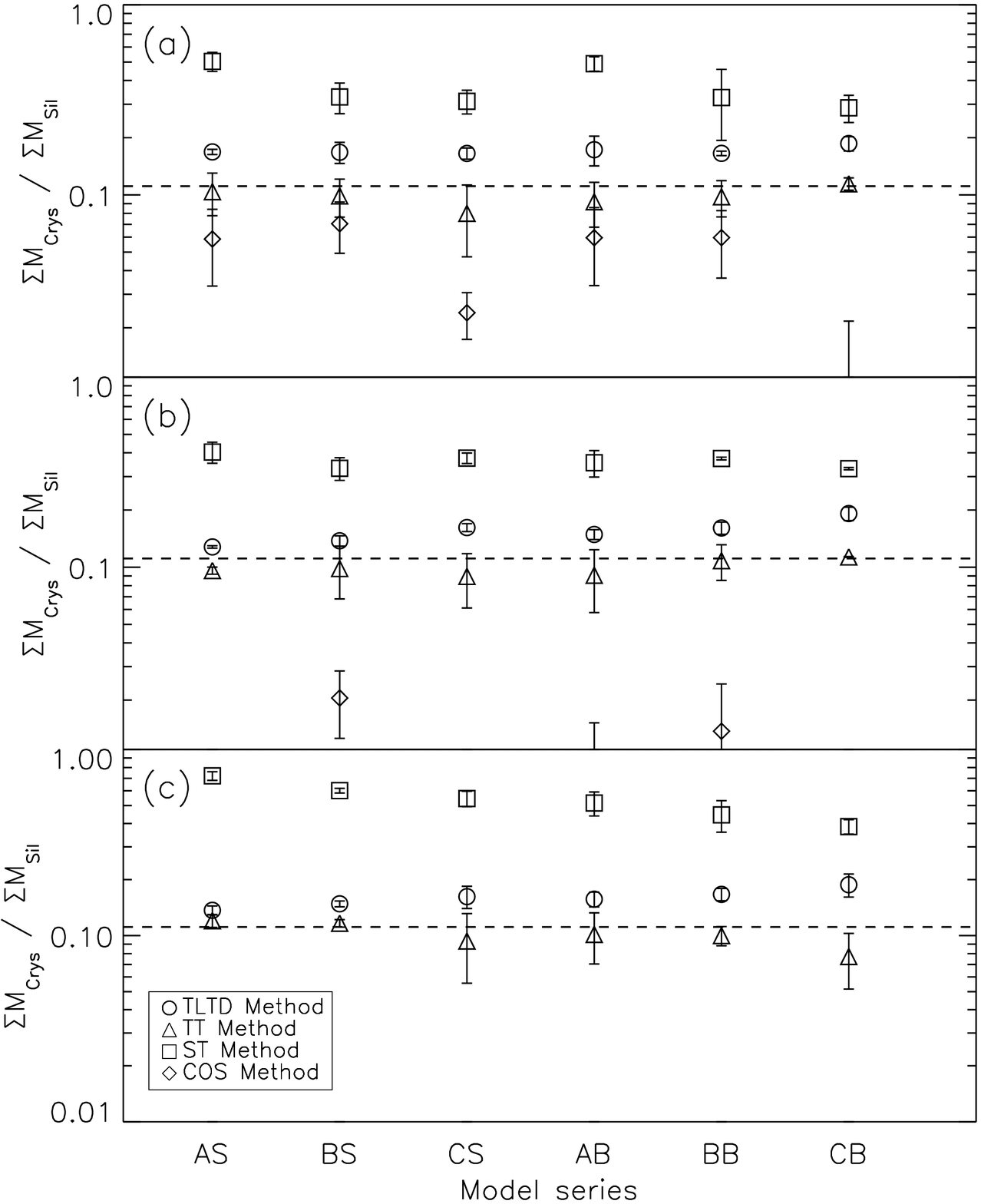}
\includegraphics[width=8.2cm]{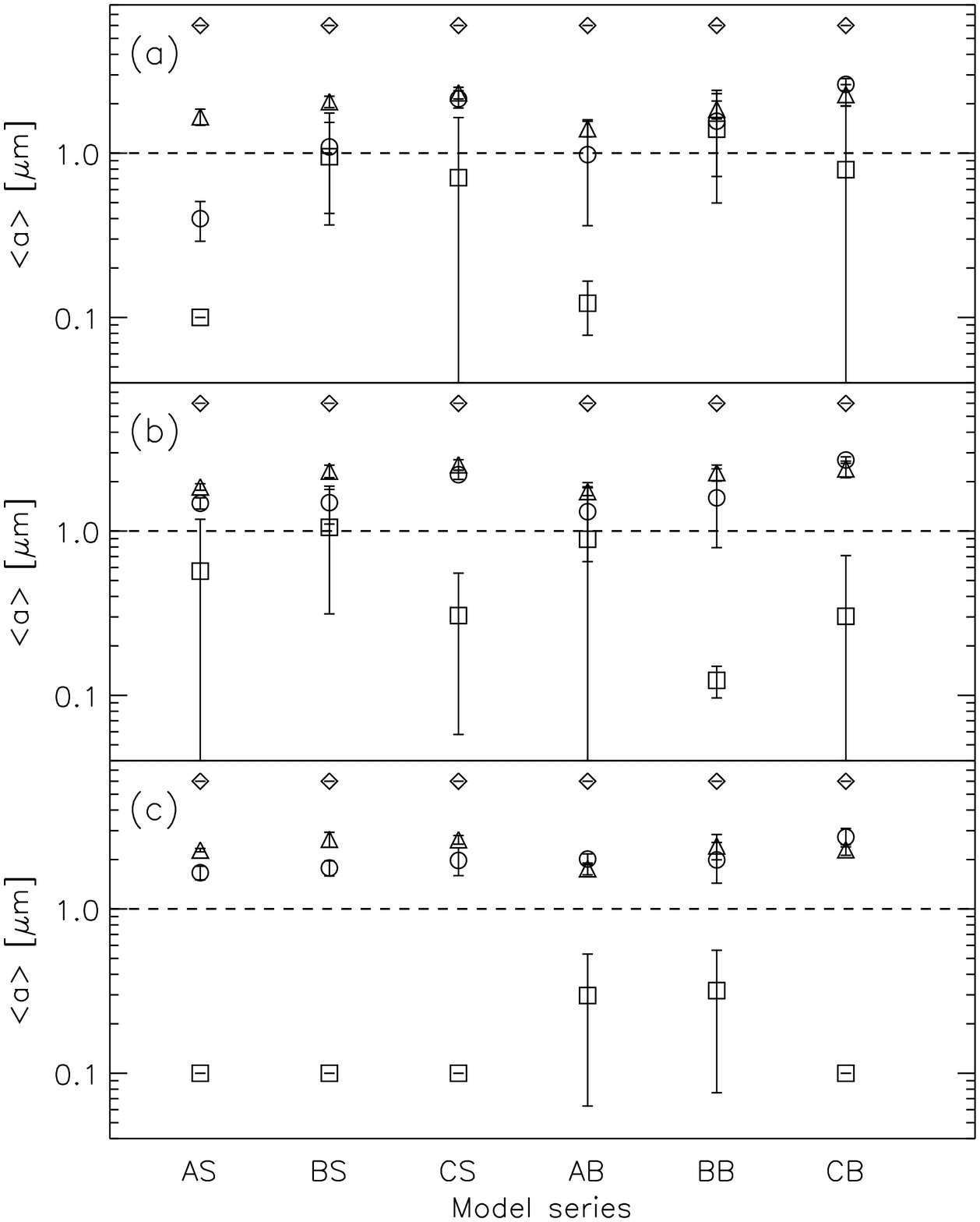}
\caption{Fraction of silicate mass in crystals ({\it Left}) and  mass-averaged
grain size ({\it Right}) for ({\it a}) Herbig Ae star, ({\it b}) T Tauri star and 
({\it c}) Brown Dwarf. The spectra were fitted without carbon among the fitted dust
species, however carbon is always present in the 2D RT models. 
The symbols show the averaged value over the different inclination angles, and the 
error bars show the standard deviation. The dashed lines show the input value to the RT code.}
\label{fig:comp_carbon_out}
\end{figure*}

In order to study the role of carbon grains in our fits, we tried to fit the same model spectra 
\emph{without} carbon (see Fig.\,\ref{fig:comp_carbon_out}), however carbon is present 
in the input dust composition \footnote{In reality carbon grains can very well be an important grain component
of protoplanetary disks.}. We want to note, that real observations were always fitted without
carbon if the COS, the ST or the TT method has been used for spectral decomposition.
In the case of the COS method the resulting crystallinity 
is completely underestimated compared to the input value (down to 0\,\% for Brown Dwarf spectra,
see Fig.\,\ref{fig:comp_carbon_out}\,{\it Left}). 
Moreover, the COS method fits lower crystallinities for lower stellar luminosities (r=0.75, p=$10^{-14}$). 
The resulting crystallinity fraction of the ST is overestimated by a factor of 4$\pm$1 compared 
to the 'real' value. The resulting crystallinity in this case depends weakly on the stellar
temperature (r=-0.4, p=$6\cdot 10^{-4}$) and on the flaring index (r=-0.43, p=$1.5\cdot 10^{-4}$)
and the fitted crystallinity is higher compared to the fits with carbon. 
The TT method underestimates the crystallinity slightly (by a factor of 0.85$\pm$0.2). 
The results are sensitive to the inclination of the disk, indicated by the larger error bars in 
Fig.\,\ref{fig:comp_carbon_out}\,{\it Left}. The results do not correlate with the flaring index, 
or stellar temperature but depend weakly on the inclination angle (r=0.43, p=$1.5\cdot 10^{-4}$). 
The average crystallinity evaluated by the TLTD method is overestimated by 
a factor of 1.5$\pm$0.4 on average.
The normalized crystallinity fitted by the TLTD method is less sensitive to the inclusion 
or exclusion of featureless grains compared to other methods.

The mass-averaged grain size of the amorphous silicates is presented in 
Fig.\,\ref{fig:comp_carbon_out}\,{\it Right} for fits without carbon among the fitted dust species.
The exclusion of the featureless dust species results
in higher average grain sizes for the silicate species for the COS and TT methods by
a factor of 2 and 1.4, respectively, compared to the fits with carbon. There is a strong 
inverse correlation between the average grain size, predicted by the TT method, and the
stellar temperature (r=-0.42, p=$2\cdot 10^{-4}$) and the flaring index (r=-0.66, p=$2\cdot 10^{-10}$). 
In the case of the ST method the resulting mass-averaged grain size of the amorphous silicates decreases 
slightly (by 20\,\%) compared to the fits with carbon. The average grain size estimated by the TLTD
method decreases on average compared to fits with carbon by 20\,\%. Furthermore, the
the predicted grain size correlates with the flaring index (r=-0.49, p=$8\cdot 10^{-6}$).

\subsection{Wavelength range}
\label{sec:fit_range}

The wavelength range, for ground-based observations, is usually limited to 8--13\,{\micron} due to 
the strong absorption of the Earth's atmosphere both shortward and longward of this range. For 
{\it Spitzer} data, the broadest possible wavelength range is determined by the 
instrumental capabilities of the IRS instrument (5.2--38\,{\micron}). 
It may be important how the wavelength range for the fit is chosen.
A narrower wavelength range corresponds to a narrower range of
temperatures one has to take into account in the modelling of the feature\footnote{For 
a narrower temperature range, the assumption of a fixed dust composition is certainly better
justified compared to a wide range in temperatures which corresponds to a larger disk region.}. 
On the other hand, a narrower wavelength 
range can result in a higher degeneracy among the optical data of different dust species. 
This effect results in a higher uncertainty in the derived dust parameters. 
In order to investigate how the results depend on the choice of the 
selected wavelength range, we fitted all the T Tauri spectra with different wavelength domains. 
Three wavelength regions were used for the ST and TT methods (8--13\,{\micron}, 
7--14\,{\micron}, 7--17\,{\micron}), while these ranges were supplemented by a fourth very broad
range for the TLTD method (5--35\,{\micron}). 

\begin{figure*}[!t]
\includegraphics[width=8.2cm]{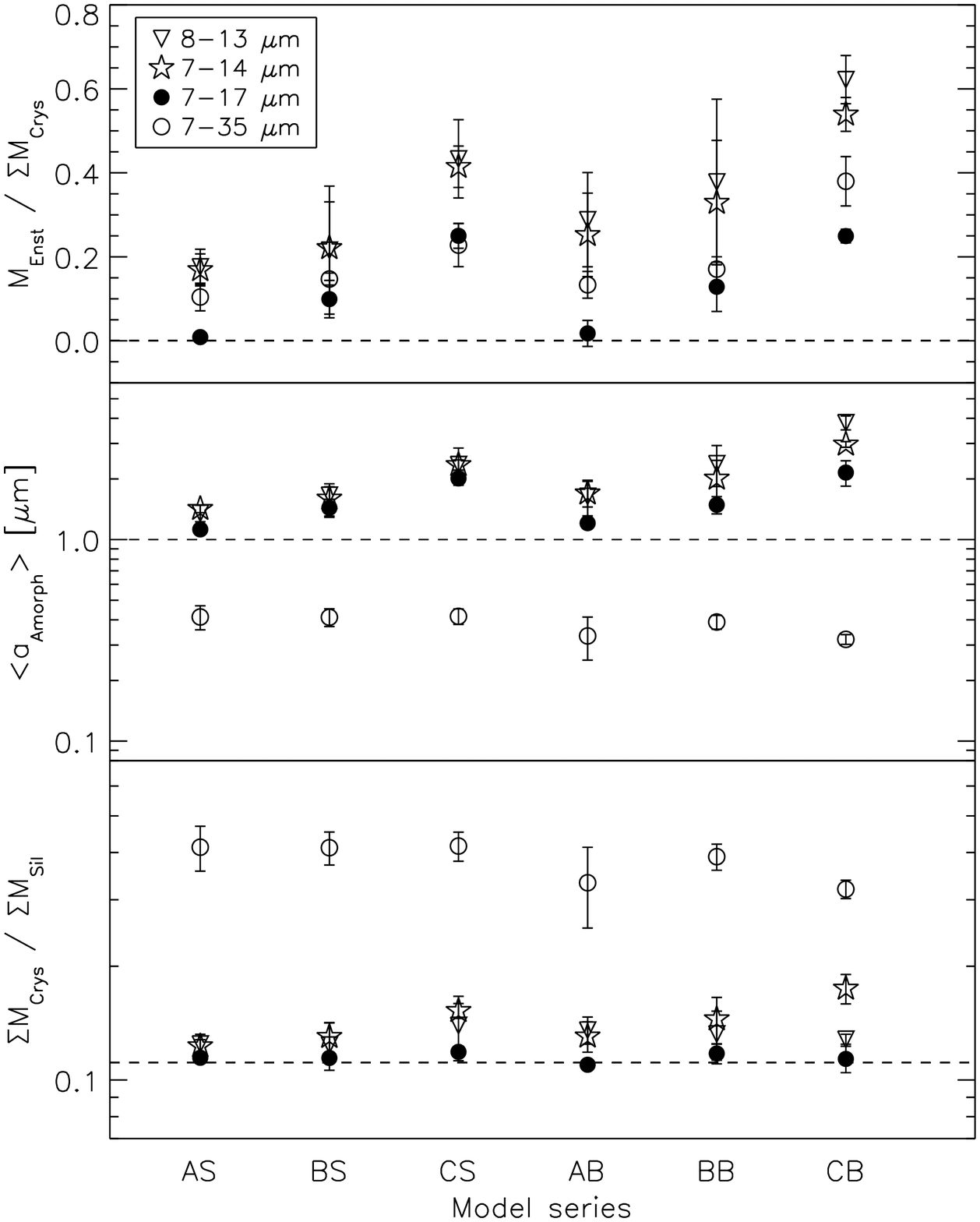}
\includegraphics[width=8.2cm]{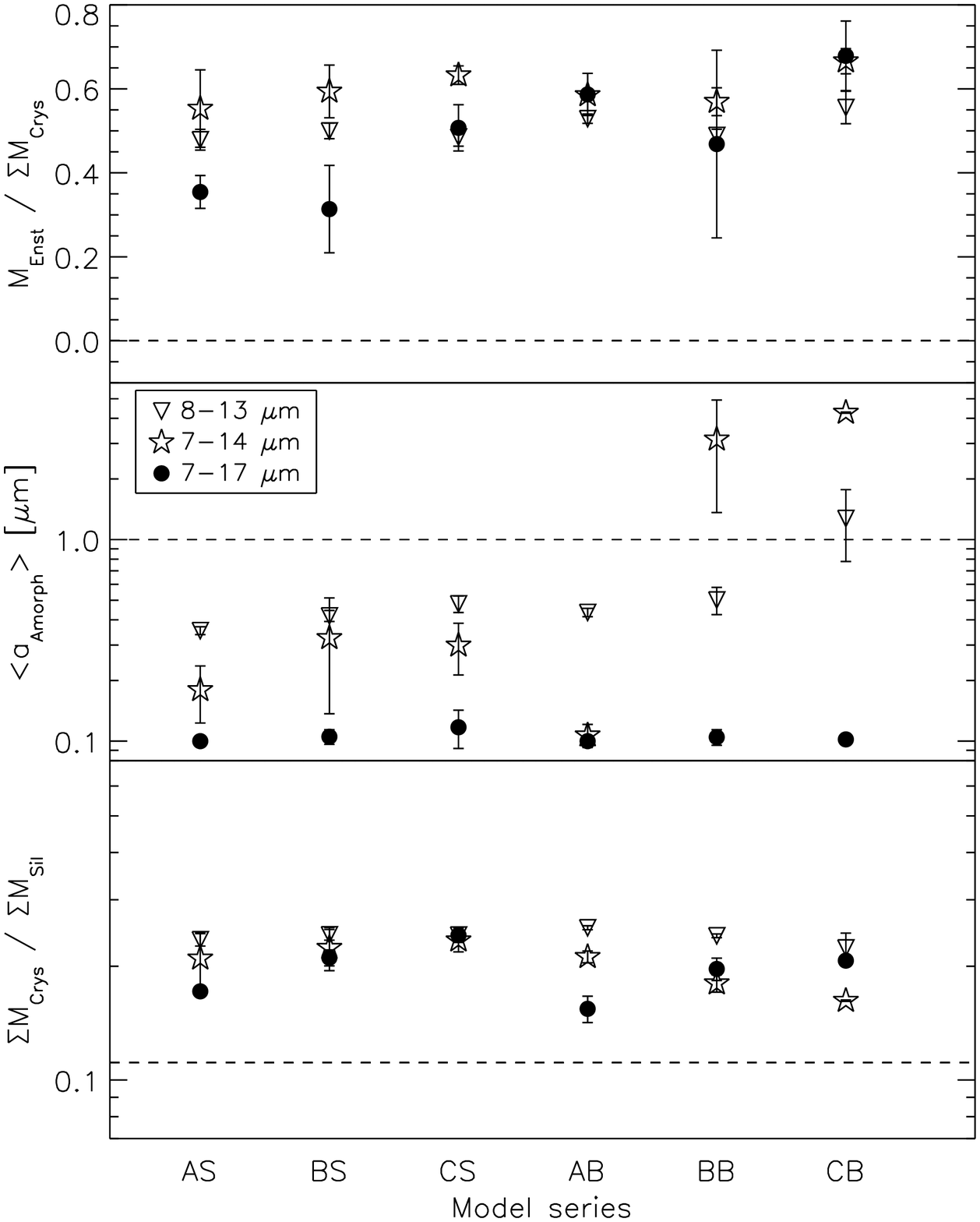}
\caption{The effect of the fitted wavelength range on the results of the T Tauri spectra, 
fitted by the TLTD method ({\it Left}) and by the TT method ({\it Right}). 
The symbols show the averaged value over the different inclination angles, and the error bars 
show the standard deviation. The dashed lines show the input value to the RT code.}
\label{fig:td_range}
\end{figure*}

The results showed that, for the COS, ST and TT method, the 7--14\,{\micron} range gives the best 
results in general. However, the estimated large enstatite particle fraction is much closer to the "real" 
values, using a broader wavelength range (see Fig.\,\ref{fig:td_range}{\it Right}). 
The estimated mass-averaged grain size of the amorphous silicates is 
underestimated with the other wavelength intervals (roughly by a factor of two) compared to the 
results of the fit using the 7--14\,{\micron} domain (see Fig.\,\ref{fig:td_range}{\it Right}). 
In the case of the TLTD method, the broader wavelength range results
in better estimates of all parameters. It is interesting to note that we get the 
best results using the 7--17\,{\micron} range and the results become worse if we use the broadest 
range between 5 and 35\,{\micron}. Using the 7--17\,{\micron} wavelength interval the 
mass-average grain size and total crystallinity, fitted by the TLTD method, are overestimated by 
by a factor of 1.45 and 1.05, respectively. In this case, the only correlation we found is the dependence 
of the mass-averaged grain size on the flaring index of the disk (r=-0.7, p=$8\cdot 10^{-12}$)
Using the broadest wavelength range, the average grain size is underestimated by 70\,\%, while
the total crystallinity is overestimated by a factor of 1.75. One reason for that can be that even the TLTD 
method is too simple to describe the temperature structure of a real disk correctly. Another reason for the 
worse results can be found in the optical data of the different silicate species. 
While the crystalline silicates have sharp features in the 20--30\,{\micron} region, the optical 
data of large amorphous silicate grains are very smooth, which makes it difficult to recognize this 
component. 

\begin{figure*}[!t]
\includegraphics[angle=90, width=8.2cm]{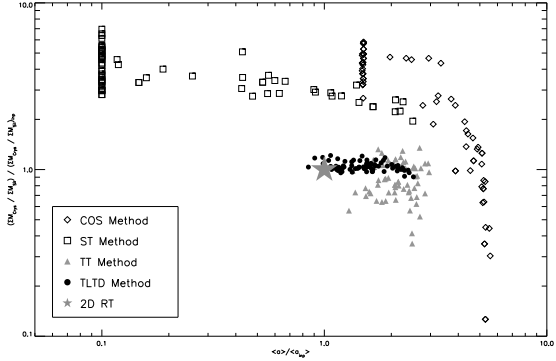}
\caption{Normalized crystallinity as a function of mass-averaged grain size for the best 
configuration of each method. A wavelength range of 7--14\,{\micron} was used
for the ST, TT and COS methods, while we used 7--17\,{\micron} for the TLTD method. The spectra
were fitted without carbon using the ST and TT methods while for the TLTD and COS methods 
carbon was included in the fit.}
\label{fig:td_range_sum}
\end{figure*}

The normalized crystallinity as a function of mass-averaged grain size is presented in 
Fig.\,\ref{fig:td_range_sum} for the best configuration of each method. The results of the TT 
method seem to be relatively close to that of the TLTD method. However, the fitted dust 
composition obtained by the TT method suffers from the overestimation of the abundance of 
large enstatite grains. The predicted amount of large enstatite grains was closer to the input value 
using a broader wavelength range (7--17\,{\micron}) regardless of the spectral decomposition method. 
This fact proved our former assumption that the overestimation of the large enstatite grains is the 
result of the degeneracy between the optical data of the large enstatite grains and other dust species in
the narrower wavelength interval. 

\subsection{Large grains}

It is well known, that the strength of the spectral features in the 10\,{\micron} region
decreases with increasing grain size (see, Fig.\,\ref{fig:opacities}). Therefore the emission
of dust grains grains larger than 2\,$\mu$m are usually neglected during the spectral decomposition.
Although the strength of the 10\,{\micron} feature of the amorphous silicates is much weaker
for a grain size of 6\,{\micron} than for 0.1\,{\micron} the feature strength is not completely
zero. In this study, we included also 6.0\,{\micron} sized grains in the input dust composition
to the 2D RT code. Dust grains with such size are probably present in the disk atmosphere if
grain growth occurs in the disk, although their abundances depend on the turbulence and
sedimentation processes.  In order to test the importance of the largest grain population
(in case of ground-based observations), all T Tauri spectra were fitted using a wavelength range of 8--13\,{\micron} 
and excluding all 6\,{\micron} sized grains from the fit, although large grains were present
in the 2D RT disk models. In this case we were interested how well the properties of the smallest
grain populations can be recovered by the methods. Therefore, during the comparison of the 
fitted dust mixtures with the input composition we used only the 0.1{\micron}  and 1.5{\micron} 
sized grains, which is the usual approach in the literature.

\begin{figure*}[!t]
\includegraphics[angle=90, width=8.2cm]{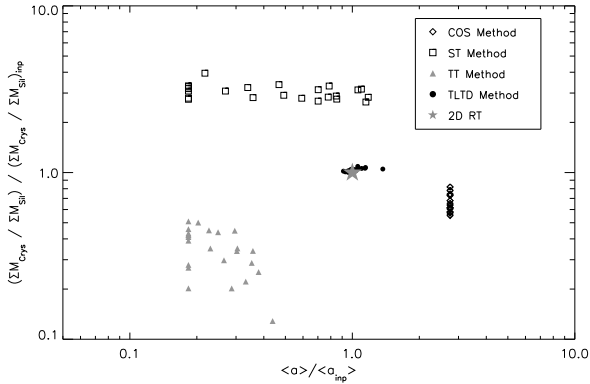}
\caption{Normalized crystallinity as a function of mass-averaged grain size for all T Tauri spectra
fitted without 6\,{\micron} size grains. A wavelength range of 8--13\,{\micron} was used
for all methods. Carbon was included in the fit only in the case of the TLTD method.}
\label{fig:td_range_sum_sg}
\end{figure*}

The results are summarized in Fig.\,\ref{fig:td_range_sum_sg}. It can be easily seen that
the dust parameters fitted by \emph{all} methods deviates from the input dust parameters.
If we included large grains in the fit, their estimated fraction by the ST method was less than a few percent, therefore
their effect on the fit is obviously weak. The resulting dust parameters of the TT method,
get further away from the right values if we excluded the large grains from the fit. The smaller
mass-averaged grain size is caused by the excluded large amorphous grains, the fraction of
which is significant if we included them in the fit. The smaller estimated crystallinity is
a result of the exclusion of large enstatite grains. The results of the TLTD method were almost
identical to the input dust mixture giving far the best results among the tested methods.
The differences between the resulting dust composition including/excluding large grains in/from the fit
can be related to the disk structure and will be discussed in detail in Sec.\,\ref{sec:discussion:diskstruct}.

\subsection{The influence of simulated noise}
\label{sec:noise_level}

Observations always suffer from uncertainty in the observed quantities and these errors naturally
affect the parameters derived from the observational data. Since one cannot avoid the effect of
noise, it is important to know how sensitive the applied method is to the noise level and
how the uncertainty of the derived parameters can be estimated from the observational errors. 
In the case of the spectral decomposition of the silicate feature around 10\,{\micron}, 
it is very difficult to describe the error propagation by analytical formulas. A simpler way to  
estimate the uncertainty in the fitted chemical composition is a Monte Carlo type of error estimation
(e.g.,\ \citealt{ref:van_boekel_2005a}).
In this kind of error estimation, a normally distributed noise is added to the spectrum, scaling 
the width of the distribution to the simulated observational uncertainty in the flux value.
Then the resulting spectrum should be fitted. This procedure should be repeated many times. Then 
the standard deviation of the resulting mass fractions will be the uncertainty of the derived dust  
compositions. 

We used the spectrum of a T Tauri star with a moderately flared disk and an inclination of 
45$^\circ$ for testing the effect of different noise levels on the results of the TLTD method.
We studied the effect of noise on ground- and space-based observations, using a wavelength range
of 8--13\,{\micron} and 7--17\,{\micron}, respectively.
Three different noise levels have been used in these tests with an $F_{\nu}^{\rm{error}}/F_{\nu}^{\rm{observed}}$
of 0.1, 0.01 and 0.001, corresponding to a signal-to-noise (S/N) ratio of 
10, 100 and 1000, respectively. We want to note, that the noise level was assumed to be independent
of wavelength, which can be good approximation for bright sources, while this assumption is certainly
not correct for faint sources with strong silicate feature.
After 100 spectra have been generated and fitted for each noise level, we studied the average 
chemical composition over the 100 spectra, the mass-averaged grain size and normalized crystallinity 
and the scattering around the average values. The latter two parameters have the advantage that, even for low S/N, 
they can be more certainly evaluated than the mass fraction of a single dust component. 

\begin{figure*}[!t]
\includegraphics[angle=90, width=16.4cm]{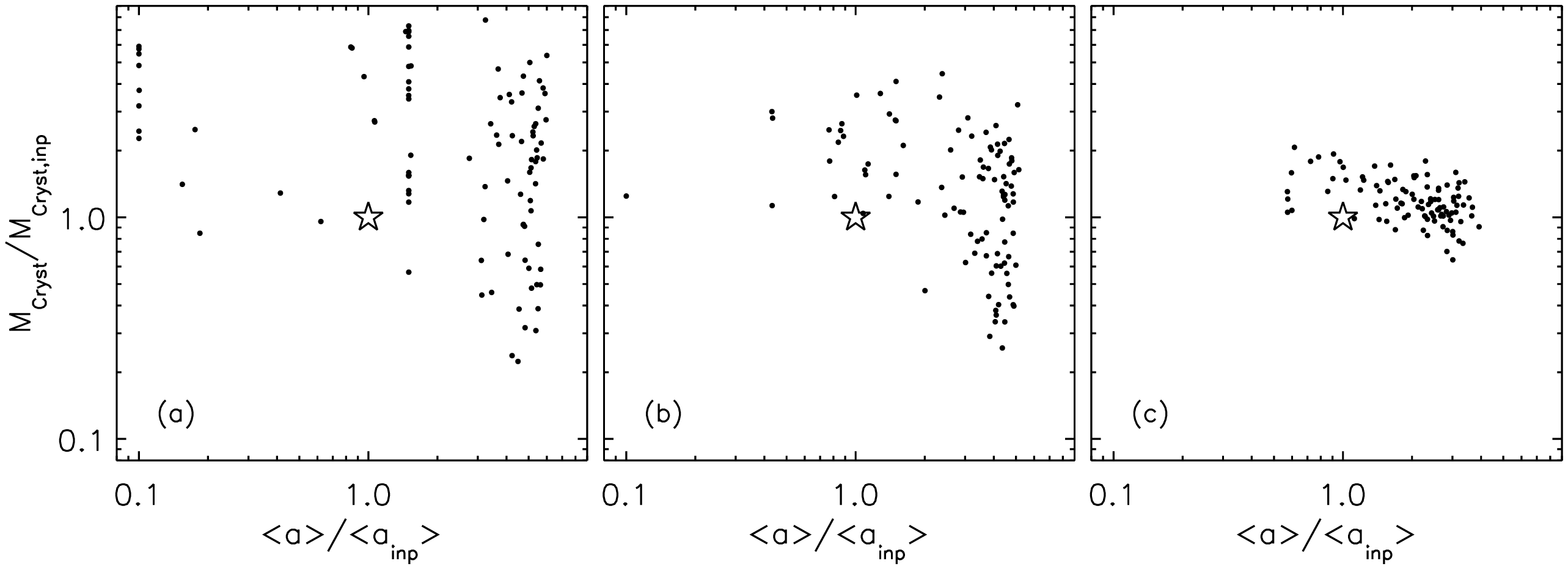}
\caption{Mass-averaged grain size as a function of total crystallinity derived with the TLTD  method. 
Both statistics are normalized to the input value of the 2D RT code (indicated by the star in the middle of the 
figures). The fitted wavelength range was 8--13\,{\micron}, while the applied signal to noise 
ratios were {\it a}), 10, {\it b}) 100, {\it c}) 1000.}
\label{fig:noise_8_13}
\end{figure*}

The normalized crystallinity as a function of mass-averaged grain size, derived by the TLTD 
method, is presented in Fig.\,\ref{fig:noise_8_13} for simulated ground-based observations 
(8--13\,{\micron}). It is easy to see , that the uncertainty in the value of crystallinity and mass-averaged 
grain size (obviously) increases with increasing noise level. Moreover, for lower signal-to-noise
ratio the average value of the crystallinity and the average grain size increases.  For a S/N of 100 the
average value of the crystallinity and the mass-averaged grain size are overestimated by a factor of 1.5
and 3, respectively, compared to the input values.  The scattering of the individual points is not perfectly 
symmetric to the mean value. The uncertainty of the value of the crystallinity is higher above the
mean value than below that. The uncertainty on the value of crystallinity for a S/N of 100 is +11\,\% -7\,\%
in an \emph{absolute} sense (i.e. in terms of the total silicate mass). 
In the case of the average grain size the uncertainty on the average value is +1\,$\mu$m and -1.5\,$\mu$m.

\begin{figure*}[!t]
\includegraphics[angle=90, width=16.4cm]{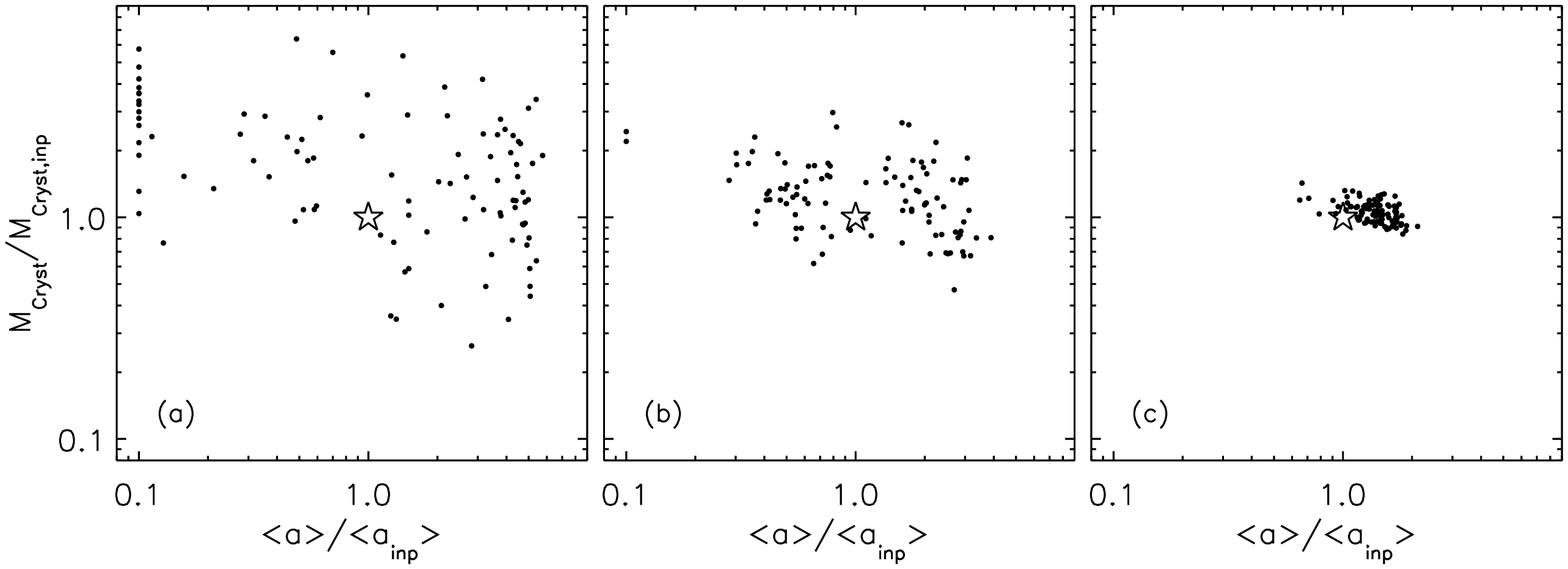}
\caption{Mass-averaged grain size as a function of total crystallinity derived with the TLTD method. 
Both statistics are normalized to the input value of the 2D RT code (indicated by the star in the middle of the 
figures). The fitted wavelength range was 7--17\,{\micron}, while the applied signal to noise 
ratios were {\it a}) 10, {\it b}) 100, {\it c}) 1000.}
\label{fig:noise_7_17}
\end{figure*}

 For space-based observations (see Fig.\,\ref{fig:noise_7_17}, 7--17\,{\micron} range), the scattering of the normalized 
crystallinity and  mass-averaged grain size decreases compared to that of the ground-based 
observations. Furthermore the average values of these quantities are closer to the input value 
of the 2D RT codes using the broader wavelength range. For a S/N of 100 the average value of the crystallinity and the
mass-averaged grain size are higher than the input value by a factor of 1.26 and 1.44, respectively. The uncertainties
on the average value of the crystallinity is 5\,\% in an \emph{absolute} sense (i.e. in terms of the total silicate mass).
In the case of the average grain size the uncertainty on the average value is 0.8\,$\mu$m.

Although the signal-to-noise ratio of the data is of fundamental importance, the uncertainty of the 
fitted dust parameters, caused by the noise, depends on the actual dust composition 
and the optical constants. Since the crystalline silicates have usually strong, narrow and sharp 
features their abundances can be constrained better, even for low signal-to-noise ratio, than the 
abundances of the amorphous silicates with broad, flat features. Therefore, it is obvious, that the 
higher the crystallinity the easier it is to determine its value.

\section{Discussion}
\label{sec:discussion}

Our results clearly indicate that the evaluation of the chemical composition of dust in 
protoplanetary disks from mid-infrared spectra is not straightforward 
even if do not have to consider the uncertainties in the optical constants of the dust grains. 
The resulting dust parameters strongly
depend on the assumed continuum below the feature and the assumed temperature(s) 
of dust grains emitting at the considered wavelengths. In the following, we will discuss 
the reasons why the results of different methods deviate from the input dust composition, 
and which effects should be taken into account during the analysis of mid-infrared dust
features. 

\subsection{Temperatures in the feature}
\label{sec:discussion:tfeat}

\begin{figure*}[!t]
\includegraphics[width=8.2cm]{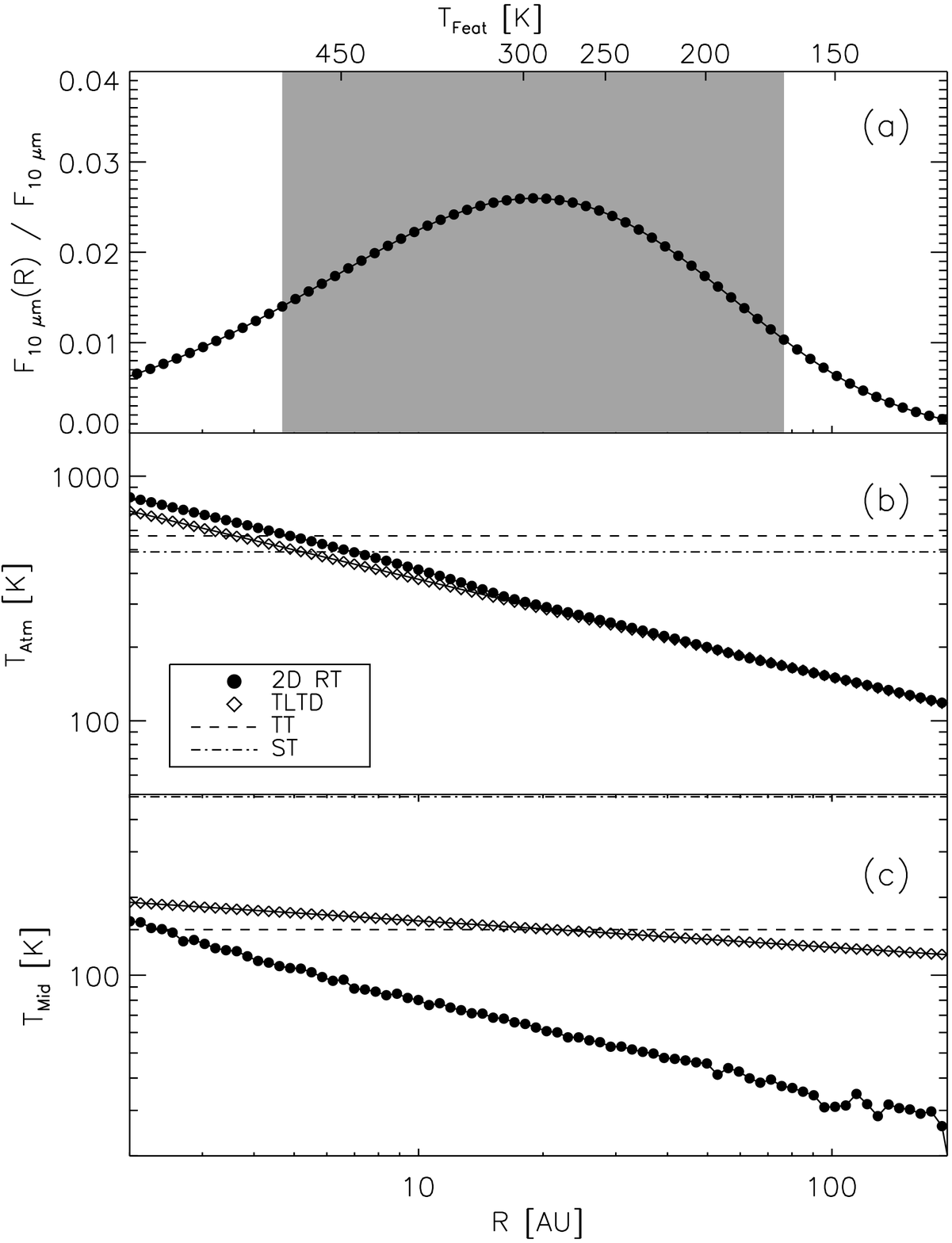}
\includegraphics[width=8.2cm]{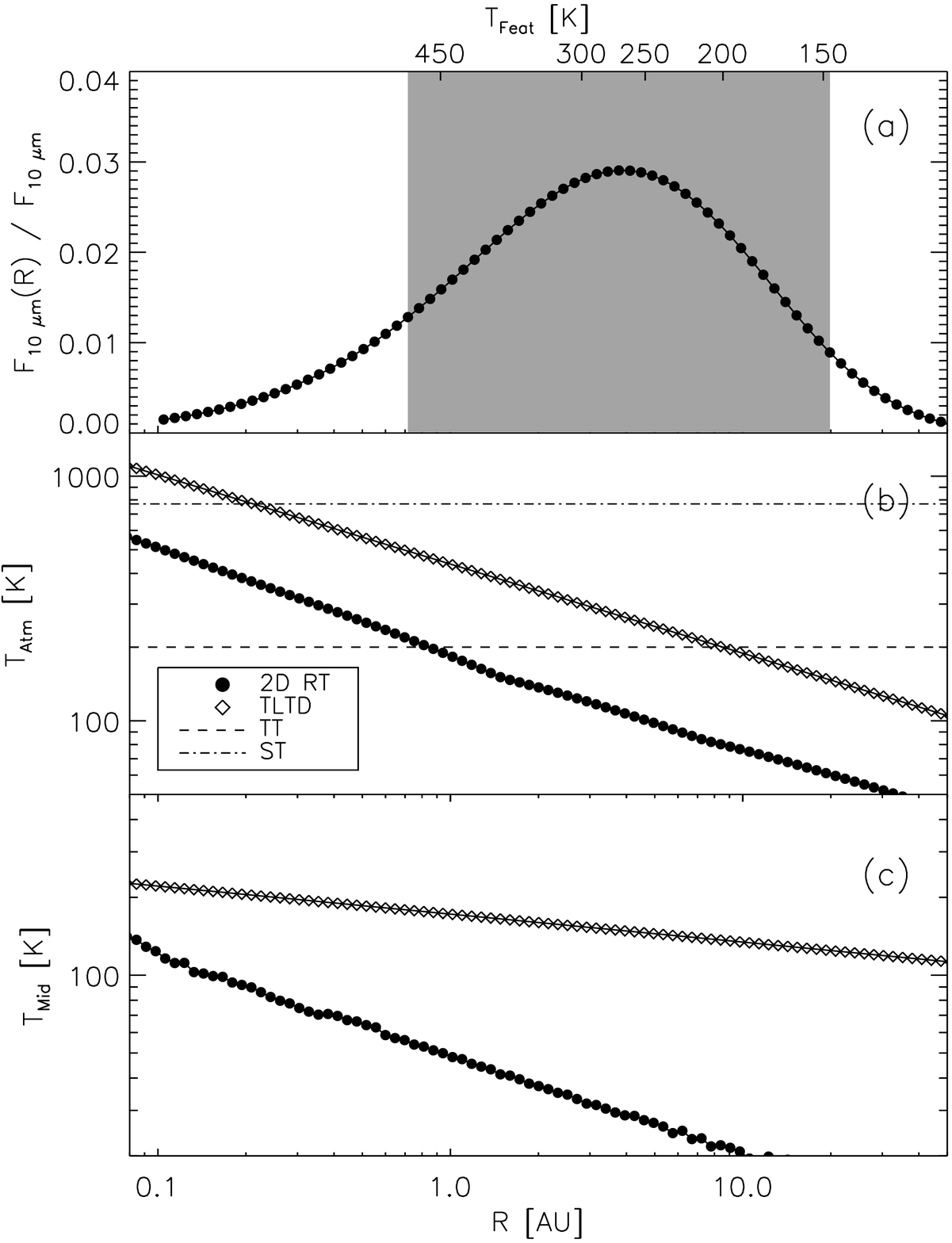}
\caption{(a): Contribution of different annuli to the total flux at 10\,{\micron} in the
TLTD fit of a Herbig Ae disk model ({\it Left}) and a Brown Dwarf disk ({\it Right}) from the BS model
series at an inclination angle of $45^{\circ}$. The grey area marks the region from which 70\,\% 
of the total flux comes from. 
(b): Radial distribution of the temperature in the disk atmosphere. It has
been calculated for the 2D RT model (filled circles) , the results of the TLTD code 
(hollow diamonds), the result of the TT method (dashed line) and the ST method 
(dot-dashed line). (c): Radial distribution of the temperature in the continuum. 
The symbols are the same as in the figure in the middle.}
\label{fig:cumflux} 
\end{figure*}

Both the ST and TT methods assume that \emph{the source function underlying the emission in the 10\,{\micron} feature can be 
well approximated by a blackbody curve with a single effective temperature}. \citet{ref:van_boekel_2005b} have shown 
that the region where the 10\,{\micron} emission comes from can extend from less than 1\,AU 
to several tens of AU in Herbig  Ae systems. They also have shown that the mid-infrared radiation, 
emitted by the different annuli (with different temperatures), contributes to the total 
observed flux in this wavelength regime significantly. 
We also investigated the temperature structure of the 2D RT disk models in the region where 
the bulk of the 10\,{\micron} emission originates. Images of the disk models were calculated 
at 10\,{\micron} and then the cumulative flux as a function of radius was used to determine 
the size of the region from which 70\,\% of the total flux originates (Fig.\,\ref{fig:cumflux}{\it a}). 
In order to characterize the temperature in the disk atmosphere we evaluated the temperature 
along the line where the radial optical depth $\tau_{0.55\,\mu{\rm m}}$ reaches unity in the disk, using 
the definition of \citet{ref:cg97}. From Fig.\,\ref{fig:cumflux}{\it b} it can be seen, that
the temperature in the disk atmosphere changes significantly in the region where the silicate 
emission comes from. On the other hand the changes of the surface density of the disk atmosphere as a function
of radius are just marginal and they depend on the flaring index. Since the silicate emission feature
is produced by a convolution of optical depth (or surface density) and the Planck function at different
temperatures, it now becomes obvious why the simple "isothermal" models give poor results. A sum
of different Planck curves with these rather different temperatures cannot be reproduced by a single 
Planck curve. 

Since we used the same $R_{in}$ and $R_{out}$ both in the 2D RT code and in the TLTD fitting
method, we can compare the fitted temperature distribution with the temperature structure
of the 2D RT models. Fig.\,\ref{fig:cumflux}{\it b} shows that the TLTD method
can efficiently evaluate the temperature distribution in the disk atmosphere. 
The ratio of the slope of the fitted atmosphere temperature by the TLTD method
and that of the 2D RT code was 0.88$\pm$0.07 averaging over all the fitted spectra.
The same ratio for the boundary temperature ($T_{{\rm atm,max}}$) was 1.3$\pm$0.6. 

\subsection{Temperatures in the optically thick continuum}
\label{sec:discussion:tcont}

Another assumption of the ST and TT methods is that \emph{the continuum below the 
emission feature can be described by a Planck function with a single temperature}.
Using the passive irradiated disk model of \citet{ref:ddn2001}, the continuum 
emission in that region has four components: emission of the central star, the 
puffed-up inner rim, the disk midplane and the emission of the featureless (e.g.,\ carbon) 
grains in the disk atmosphere. 
In general, shortward of $\sim$8\,{\micron}, the continuum is dominated by the emission of 
the hot inner rim of the disk with decreasing flux towards longer wavelengths.
At wavelengths longer than 12-14\,{\micron} most of the continuum emission 
comes from the disk midplane with increasing flux towards longer
wavelengths. Thus, the slope of the continuum should change in the 
8--12\,{\micron} region in order to match the decreasing flux of the inner rim
and the rising flux of the midplane. The turn-over point, where the slope changes 
is determined by the balance of the rim and midplane components (which is set by 
the disk geometry and the inclination) and by the featureless grain 
(e.g.,\ carbon) content of the dust. One of the reasons for the poor performance of the 
COS, ST and TT methods lies in their incapability to fit a continuum with such a shape. 

\begin{figure*}[!ht]
\includegraphics[angle=90, width=16.4cm]{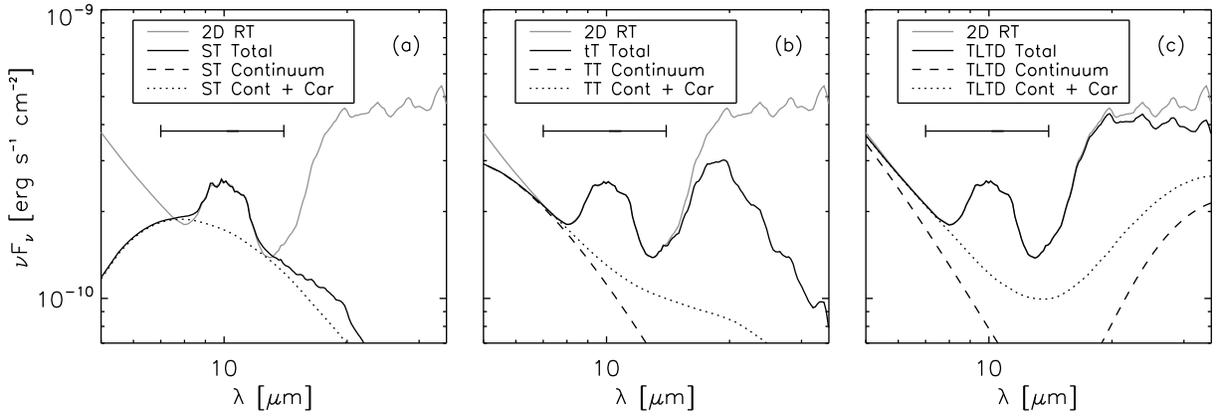}
\caption{Spectrum of a T Tauri model in the AS series at an inclination angle of $45^{\circ}$ fitted by 
{\it a}) the ST method, {\it b}) the TT method and {\it c}) the TLTD method. The horizontal
line shows the wavelength range of the fit.}
\label{fig:cont_fit}
\end{figure*}

The ST method simply cannot reproduce the real continuum, since the temperature in
the continuum and in the feature is the same in this approach (see Fig.\,\ref{fig:cont_fit}{\it a}). 
The only way to fit the real continuum is the inclusion of large amorphous carbon, 
the mass absorption coefficient of which increases with wavelength in this region. Therefore,
this method fitted an unrealistic amount of carbon ($\sim$70--90\,\% of the total mass)
in order to fit the real continuum.

\begin{figure*}[!ht]
\includegraphics[angle=90, width=8.2cm]{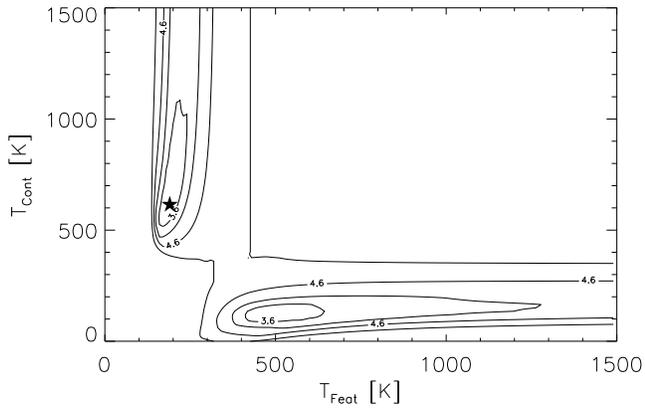}
\caption{Contours of $log(\chi^2)$ for the TT method during the fitting of the
T Tauri model in the AS series at an inclination angle of $45^{\circ}$. The black star marks the 
global minimum and the contour lines are separated by 0.5.}
\label{fig:logchi_tt}
\end{figure*}

In the case of the TT approach two distinct solutions exist for the fit, which can 
be easily seen in Fig.\,\ref{fig:logchi_tt} as the $\chi^2$ space has two minima.
This kind of $\chi^2$ space is characteristic for the TT method during the fitting
of the spectrum of a Class II source, using the classification scheme of \citet{ref:lada_1984}
and related to the fit of the continuum. One of the two minima has a high continuum 
temperature ($\sim 600$\,K) and low feature temperature ($\sim 200$\,K, see Fig.\,\ref{fig:cont_fit}{\it a}). 
In this case, the resulting dust composition overestimates the mass fraction of large 
($a=6\,\mu$m) carbon grains. In this solution the real continuum is fitted with the high
temperature continuum component shortward of $\sim$10\,{\micron}, while at longer wavelengths 
the real continuum is fitted by the optically thin emission of carbon grains.
In the case of the other minimum, the feature has a higher temperature than the continuum 
and the resulting amount of carbon is very well estimated. 
On the other hand the silicate grains, producing the 10\,{\micron} 
feature, are significantly larger compared to those in the other solution.
The reason for the larger average grain size is that the mass absorption coefficient
of the large (6\,{\micron}) amorphous grains is higher than that of the smaller ones 
shortward of $\sim$8.5\,{\micron}. Therefore the real continuum emission at longer wavelength
is fitted with the single temperature continuum of the TT method, but shortward of 8.5\,{\micron}
the real continuum is fitted by including large amorphous grains. 

The subtraction of the continuum can be a better solution if, and only if, the real
continuum is known, otherwise the subtraction of the continuum introduces an additional
uncertainty in the results. The highest uncertainty in the COS method is that it tries to 
extrapolate the continuum by fitting a few data points. Therefore, the fitted
continuum is very sensitive to the wavelength ranges in which it has been fitted. Although,
the COS method estimated the mass-averaged grain size in a few cases equally well as the TLTD method, 
the total crystallinity was deviating most strongly from the real value. 

We compared the temperature profile of the disk midplane calculated by the 2D RT code to 
the fitted temperature distribution by the TLTD method (Fig.\,\ref{fig:cumflux}{\it c}). 
In this case the temperature distribution in the 2D RT model was calculated along the line 
where the vertical optical depth in the disk at 10\,{\micron} equals unity. One can see that 
the TLTD method overestimated the temperature compared to that of the 2D RT code. Furthermore 
the slope of the temperature distribution is very low. The ratio between the fitted power-law 
indices and the calculated one by the 2D RT code was 0.33$\pm$0.33, and that of the
boundary temperature was 1.4$\pm$1.0. The reason for the difference between the fitted and the 
'real' midplane temperature in general is caused by the fat that even the TLTD method is too
simple to represent the temperature structure of a protoplanetary disk. In contrast to the 7--8\,$\mu$m region,
longward of 12\,$\mu$m the continuum cannot be unambiguously determined, since the
dominant contributor to the total emission is still the disk atmosphere. Moreover, there is a degeneracy
between the optically thin emission of the featureless grains and the optically thick disk emission.
In a 2D RT code the layer where the 'midplane' emission originates depends strongly on the 
disk geometry, inclination angle, and of course on the wavelength. During the comparison of the
results of the TLTD method and the 2D RT code we made the simplifying assumption that the
layer, where the 'midplane' emission comes from, is where the vertical optical depth at 10\,$\mu$m equals unity,
which is only true for face-on orientation.

\subsection{Explanation of the correlations}
\label{sec:discussion:diskstruct}

The shape and strength of the silicate feature around 10\,{\micron} is affected by
the disk structure, the inclination and the stellar type in a complex way.
The efficiency of a spectral decomposition method depends on how it can handle these effects.
In the former two sections we discussed why the single and two temperature approximations or
a continuum subtraction are insufficient to model the temperature structure of the silicate
emission zone in a protoplanetary disk. This weakness of the fitting routines can lead to a 
large scatter in the derived dust parameters. Here, we discuss how and why the 
derived dust parameters can to depend on the geometry of the source or the stellar type. 
If such an "artificial" dependence is not properly understood, the comparison of grain parameters
derived for different source types can be misleading. 

\begin{figure*}[!t]
\includegraphics[angle=90, width=16.4cm]{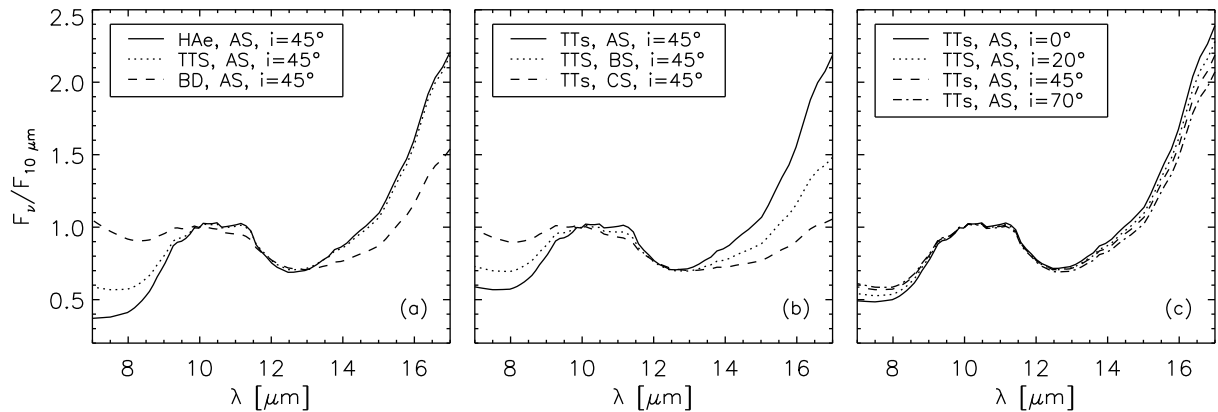}
\caption{Effects of different parameters on the 10\,{\micron} features. ({\it a}) Spectra of
flared disks around different types of stars at an inclination angle of 45$^\circ$. ({\it b})
Spectra of T Tauri stars with different disk geometries (flared, moderately flared and flat) at
an inclination angle of 45$^\circ$. Spectra of T Tauri stars with a flared disk but at different
inclination angles.}
\label{fig:explain}
\end{figure*}

The effect of the stellar temperature on the mid-infrared silicate features can be much stronger
than it is usually thought. Under the assumption of an identical star-to-disk luminosity ratio and
disk structure, the 
contribution of a Brown Dwarf (T$_{\rm eff}$=2500\,K) to the 10\,{\micron} emission is roughly 
44 times higher than that of a Herbig Ae star (T$_{\rm eff}$=9500\,K). In Fig.\,\ref{fig:explain}{\it a} 
it can be seen that this increasing "stellar" contribution toward lower effective temperatures results in 
lower peak over continuum ratios of the 10\,{\micron} feature \emph{without} changing the parameters
of the dust model. 
This phenomenon naturally explains the strong inverse correlation between stellar temperature and 
the temperature fitted by the ST method (r=-0.84) and the continuum temperature fitted by the TT 
method (r=-0.92). Moreover, the feature temperature over continuum temperature ratio predicted by 
the TT method directly correlates with the stellar temperature (r=0.93). The continuum temperature
is always higher than the feature temperature for Brown Dwarfs and mostly for T Tauri stars. 
This kind of solution seems to be unphysical (compared with the radiative transfer solution of the
problem \citep{ref:cg97, ref:menshchikov_1997}). The TT method is not a solution of the radiative 
transfer problem and simply fits the dominant continuum component. As the stellar 
contribution becomes the dominant continuum component for lower stellar temperatures, the TT method 
fits higher continuum temperatures. This behaviour of the TT method leads to a strong "artificial" inverse correlation
between the stellar temperature and the mass-averaged grain size of the amorphous silicates if we
fit the spectra without carbon. The TLTD method is much less affected by this effect, since the
stellar contribution is already included in the fit. The average grain size does not depend
on the stellar temperature if we include carbon in the fit, but it does if we exclude carbon, although
this correlation is weaker compared to that of the other methods (r=-0.67).

The effect of the flaring index, i.e. the disk geometry, is more complex than that of the 
stellar temperature. As \citet{ref:ks_2007} have shown the radius of the zone where the 10\,{\micron}
emission comes from can decrease for flattened disks compared to flared disks.
Thus the decrease of the flaring index can have two important consequences concerning the 10\,{\micron} 
silicate emission band. One effect is that the contribution of the midplane component to the total flux 
below the silicate feature decreases toward flattened disks \citep{ref:DD04}. In other words, the dominant 
continuum temperature shifts toward higher values toward lower flaring indices. Moreover, the amount of 
optically thin matter decreases more rapidly with radius in flattened disks than in flared disks \citep{ref:ks_2007}.
This behaviour leads to the decrease of the height of silicate emission feature above the continuum 
toward flattened disks (see Fig.\,\ref{fig:explain}{\it b}). The 10\,{\micron} feature is, however, less 
affected by the disk geometry for Brown Dwarfs compared to Herbig stars, since the relative contribution 
of the "stellar" component to the continuum emission at 10\,{\micron} becomes stronger for lower stellar 
temperatures. This behaviour explains why the temperature fitted by the ST method inversely correlates with the
flaring index for Herbig Ae spectra (r=-0.85) but not for Brown Dwarf spectra (r=-0.33). The continuum
temperature as well as the continuum over feature temperature ratio predicted by the TT method behaves 
exactly the same. This is likely the reason why the mass-averaged grain size inversely correlates
with the stellar temperature for all the methods (however the strength of the correlation depends on
the method), but this correlation is the strongest for Herbig Ae spectra and almost disappears for Brown 
Dwarf spectra. The TT method is an exception as the predicted grain size correlates inversely with the 
flaring index also for Brown Dwarf spectra. 

The results of none of the methods show a close correlation with the inclination angle. The 
scatter in the derived dust parameters is the smallest for the TLTD method (more than a factor of 
two compared to other methods). It can be seen in Fig.\,\ref{fig:explain}{\it c}) that the changes in 
the inclination angle modifies the relative contribution of different (mainly continuum) components 
(i.e., star, inner rim, 
and disk midplane) to the total flux. The high temperature continuum components (the star and the inner 
rim) becomes stronger for higher inclination angles (close to edge-on) while at the same time the 
midplane component becomes weaker. The TLTD method can better handle this behaviour due to the $C_{0}$,
$C_{1}$, $C_{2}$ factors. We want to note, that the TLTD method, in this form, cannot handle edge-on disks 
where the silicate emission band is affected by the extinction of the outer parts of the disk.

In general, all these findings should serve as a warning, that correlations between grain parameters and disk/stellar
properties can be artificially produced by limitations of the spectral decomposition methods.  

\subsection{Degeneracies in the optical data}
\label{sec:degeneracies}
Another problem in the spectral decomposition of the mid-infrared silicate feature
can be the degeneracy among the optical data of the fitted dust species.  If the mass absorption
coefficients of two dust species are very similar to each other it will be hard to identify them separately
in the observed spectrum regardless of the applied spectral decomposition method\footnote{Dust 
components  will not behave the same in very wide wavelength regions. Radiative transfer calculations would 
then require to obtain dust temperatures which would allow to distinguish between the components.}.

\begin{figure*}[!t]
\includegraphics[width=8.2cm]{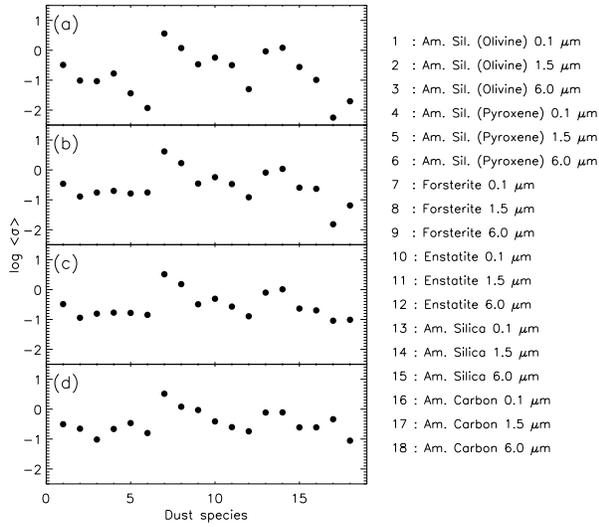}
\caption{$\sigma$ (for the definition, see Sec.\,\ref{sec:degeneracies}) as a quality indicator of how well the optical data of a specific dust species can 
be reproduced by the rest. The $\kappa_\nu$ curve of each dust species was fitted using all the other
dust species in the wavelength range of  ({\it a}) 8--13\,{\micron},  ({\it b}) 7--14\,{\micron}, ({\it c}) 7--17\,{\micron} 
and  ({\it d}) 5--35\,{\micron} . }
\label{fig:lindep}
\end{figure*}

Similar to \citet{ref:van_boekel_2005a}, we also investigated this effect by fitting the mass absorption 
coefficients of each dust species with that of the other species. In order to quantify the
quality of the fit we calculated the average deviation of the best fit model from the
mass absorption coefficient of the analyzed dust species ($<\sigma_\nu>$), where
\begin{equation}
\sigma_\nu = \left(\frac{\left(\kappa_\nu^{\rm model}-\kappa_\nu\right)^2}
{(\kappa_\nu)^2}\right)^{1/2}.
\label{eq:residual}
\end{equation}
Here, $\kappa_\nu$ is the mass absorption coefficient of the dust species under consideration, 
and $\kappa_\nu^{\rm model}$ is the mass absorption coefficient of the best fit model. 
The $\sigma_\nu$ value for each fit is presented in Fig.\,\ref{fig:lindep}.

It can be seen, that for a wavelength range of 8--13\,{\micron} the optical data of pyroxene-type
amorphous silicates and large (6\,{\micron}) enstatite can be well reproduced
by the other dust species. For this wavelength range there are several dust species for which the  
$<\sigma>$ value is less than 0.1, implying that they can be on average reproduced at a 10\,\% level by 
a linear combination of other dust species. If one has a signal-to-noise ratio of 10 these dust
species cannot be identified in the spectrum due to the similarities of the mass absorption coefficients. 
It can be seen from Fig.\,\ref{fig:lindep}\,{\it b} and Fig.\,\ref{fig:lindep}\,{\it c} that
$<\sigma>$ increases with the width of the wavelength range used for the fit. 
On the other hand, the comparison of Fig.\,\ref{fig:lindep}\,{\it c} and
Fig.\,\ref{fig:lindep}\,{\it d} shows that, for a wavelength range of 5--35\,{\micron},
the optical data of a specific dust species can be equally well reproduced by the other species,
as for a wavelength range of 7--17\,{\micron}. If the mass absorption coefficients of two dust species
are very similar to each other, then they cannot be identified in the spectrum separately 
\emph{independently from the applied spectral decomposition method}. We recommend to 
perform this kind of analysis for any new dust model, in order to recognize and understand
the possible degeneracies.

During this analysis we did not take into account the continuum since the shape of the fitted
continuum depends on the applied spectral decomposition method and we were interested in the 
degeneracies among the mass absorption coefficients only. The applied continuum has probably
the strongest effect on the featureless dust component (in our case on the carbon). Although the 
real continuum (the shape of which is the best reproduced by the TLTD method) is not a straight line
as the mass absorption coefficient of the 6\,$\mu$m sized amorphous carbon grains in the fitted domain, 
it can still be very difficult to identify such featureless dust species, especially using a narrow wavelength range 
(8--13\,$\mu$m). The mass absorption coefficient curve of the 0.1 and 1.5\,$\mu$m sized amorphous 
carbon grains can also be hardly separated from the combination of inner rim and disk midplane emission.  
Therefore, one should be careful during the interpretation of the carbon fraction fitted by simple
methods like those ones tested in this paper. 

In order to test the significance of the detection of a given dust component an F-test
can be used (e.g. \citealt{ref:min_2007}). Synthetic observations should be created by adding simulated noise to 
the observed spectrum as it has been done in Sec\,\ref{sec:noise_level}. These synthetic spectra should be fitted with 
and without the dust component under consideration. The ratio of the two $\chi^2$ distribution (with and without
the dust component) gives the F-distribution. After the mean value and the standard deviation ($\sigma$) of the F-distribution
has been calculated one can determine the distance of the mean (in terms of $\sigma$) from unity, which gives the 
significance level for the detection of the dust component. If a component, which is significantly present in the spectrum, 
is added to the fit the $\chi^2$ should decrease, while for an insignificant detection the $\chi^2$ does not change. 
One should, however, be careful with the F-test during the significance test, since the F-test assumes that all 
components are independent from each other, i.e. there is no degeneracy among the components. Without any knowledge
on the degeneracies among the dust components one should perform an F-test not only for each dust species but
also for all possible linear combination of them in order to draw the right conclusion from the test. Another
possibility can be that before the F-test is done one performs the above described simple analysis of the mass absorption coefficients
(fitting the mass absorption coefficient of a given dust species with a linear combination of others).
Such an analysis gives an indication for which dust species an F-test should be done carefully.

The general performance of a spectral decomposition method is determined by two
 important effects, which must be handled together and which act in opposite directions. 
One problem is the degeneracy/similarities of the optical data of the fitted dust species, 
which requires the broadest possible wavelength range. The other effect is linked to the fitted 
temperature(s). All the simplifying assumptions of the above described simple methods, as for
instance the assumption of an average temperature, becomes more valid for a narrower wavelength 
interval (i.e., a narrower range of temperatures), compared to a broad one. 
The optimal wavelength range for a specific method is therefore determined by the balance of
these two effects. Due to the assumption of a single temperature, all previously used methods can 
handle just a limited wavelength range (8--13\,{\micron} or 7--14\,{\micron}), where the similarities 
of the optical data can play an important role for low S/N data. An advantage of the TLTD method over the
other methods is, that one can extend the fitted wavelength interval by applying a temperature distribution.
On the other hand, the TLTD method is still an \emph{approximate} method, which means that one
cannot apply this method to an arbitrary broad wavelength range. An optimal wavelength range
can be the interval between 7 and 17\,{\micron}. For this interval, the differences between the optical data 
of the fitted dust species are roughly as high as for a wavelength range of 5--35\,{\micron}, applying our dust
model. Moreover, the 7--17\,{\micron} range is narrow enough to apply our approximate temperature
determination and allows to make the assumption that the grain composition does not change dramatically
with radius. 

\subsection{Limitations of the TLTD method}
Although the new TLTD method returned dust parameters closest to the input dust mixture in most cases,
his method has its limitations as well as any other method based on simplifying 
assumptions. During the derivation of Eq.\,\ref{eq:fit_td_temp} we assumed that the optical depth 
in the disk atmosphere does not depend on the radial distance. In a real disk the 
optical thickness in the disk atmosphere depends on the radius. However this dependence 
is marginal and depends on the flaring index. Therefore, the assumption of constant optical thickness 
in the disk atmosphere as a function of radius seems to be a reasonable assumption. 
The above described version of the TLTD method does not contain reddening. Therefore, we do not 
suggest this form of the method for disks with extremely high inclination angles 
(edge-on orientation. Here, one should perform a correction for reddening before applying the TLTD method. 
We assumed also that the 10\,{\micron} emitting region is smaller, than the outer radius of the disk.
If this assumption is not valid, then one has to fit the outer radius as well.
Despite these limitation we suggest to use the TLTD method, since the bulk of the young stars with an age 
of few Myr, where dust processing is thought to be strongest, fit within these criteria.  

Although we tested the spectral decomposition methods on spectra of numerous disk models, there
are still untested cases. We did not include radial mixing or self-consistent sedimentation in our 
2D models and we fixed the input chemical composition in order to investigate the effect of disk 
structure on the resulting dust composition. These assumptions can have an effect on the resulting chemical 
compositions of \emph{all} methods. We are aware, that our results were derived from a limited set of 
input models, based on a common parametrization. However, since we have done the analysis for a number 
of choices of parameters of the input model, we feel confident that our conclusions are robust. 
In this paper we presented an efficient method, which can be applied to analyze
large data sets in order to obtain a first estimation of the dust composition in the 10\,{\micron} 
emitting region. The main improvement in the TLTD method is the application of a temperature distribution 
to describe the temperature structure in the disk and the application of multiple continuum components. 
Since these assumptions are a better description compared to the assumption of the other simpler
methods, we think, that the advantage of the TLTD method over the other methods may not be affected by 
taking into account radial mixing, sedimentation or different chemical compositions.

\section{Conclusions}
%
Out of the three spectral decomposition methods, tested in this study, the
the two-layer temperature distribution (TLTD) method returned dust parameters
closest to the input parameters for the calculation of the synthetic spectra.
Compared to the results of the TLTD method, the fitted dust composition by the
two temperature (TT) method showed larger deviations from the input dust composition, 
while the results of the single temperature (ST) method deviated even more. Although 
the mass-averaged grain size estimated by the continuum subtraction (COS) method was 
very close to the input value in a number of cases, 
the fitted crystallinity of these models was far away from the 'real' values.

There are several reasons why the previously used spectral decomposition methods
have limited capabilities in recovering the dust composition from the 10\,{\micron}
silicate feature. The COS method tries to reconstruct a complex continuum by fitting a 
few data points outside the 10\,{\micron} silicate feature. This can modify the shape of the 
feature introducing additional uncertainties to the analysis. The ST and TT methods assume, 
that (i) the source function underlying the emission in the 10\,{\micron} feature can be 
well approximated by a blackbody curve with a single effective temperature and 
(ii) the continuum below the feature can also be described by a Planck function with a single 
temperature. Additionally, there is a degeneracy between optically thin emission of 
featureless grains (e.g.,\ carbon) and the optically thick emission of the circumstellar 
disk. Due to this degeneracy, if we included the amorphous carbon into the fit, then the 
ST method always overestimated the fraction of carbon in the spectra, while the TT method 
also had this problem in a number of cases. The TLTD suffered less from this problem compared 
to the other methods due to the more sophisticated estimation of the optically thick emission 
of the disk. 

The exclusion of carbon from the fitted dust species resulted in a better
estimation of the total crystallinity in the case of the TT method, although the
predicted mass-averaged grain size of the amorphous silicates became too small.
For the ST method the exclusion of carbon results in an overestimation of the 
total crystallinity (up to $\sim$70\%) even more so than including carbon. Furthermore, 
the ST method predicted a higher crystallinity for lower stellar luminosities if
we excluded the featureless grains from the fit. The mass-averaged grain size of the 
amorphous silicates was somewhat larger for the TLTD method without carbon in the 
fitted dust species.

All the methods overestimated the fraction of large enstatite grains using a wavelength
range of 7--14\,{\micron}. This behavior is related to the degeneracy of the mass 
absorption coefficients of large enstatite grains and small olivine, small carbon and 
medium-sized olivine grains.  The overestimation of the large enstatite fraction decreased 
if we used a broader wavelength range (e.g.,\ 7--17\,{\micron}) for the fit, where the 
degeneracy is less pronounced. 

 Although the input dust mixture contained only 4\,\% of large grains (6\,{\micron}) in 
terms of the total mass mass, the exclusion of this grain population from the fit had an effect 
on the resulting dust mixture, using a wavelength range of 8--13\,{\micron} (ground-based 
observations). The fitted dust parameters are closer to the input mixture for the COS, ST and 
TLTD methods if the 6\,{\micron} grain population is excluded from the fit. In contrast the 
TT method returned dust parameters which differed more from the input parameters if we excluded the
largest grain population from the fit compared to the results including the 6\,{\micron} grains. 
In the case of the TLTD method, the exclusion of the large grain components modified the 
mass-averaged grain size of the amorphous dust species. The changes in the estimated
fraction of the crystalline dust species, estimated by the TLTD method, are just marginal. 
In the case of the COS, ST and TT methods the fitted crystallinity was more strongly affected by
the exclusion of the 6\,{\micron} sized grains. The reason for that was partially that
by including large grains in the fit, all methods overestimated the fraction of 6\,{\micron}
sized enstatite grains.  

Here we should note that poorly performing analysis methods can lead to artificial correlations
between stellar/disk properties and dust grain parameters. The poor performance of such 
methods in representing the disk spectral energy distribution is then compensated by 
introducing emission from non-existing grain components. 

The quality of the derived dust parameters depends very strongly on the signal-no-noise
ratio of the observed spectra. For the applied set of optical constants and using
a wavelength range of 8--13\,{\micron} one can expect an uncertainty 
of about 11\,\% (in terms of the total silicate mass) in the value of the crystallinity for a 
S/N of 100. For the same signal-to-noise ration, but using a wavelength interval of 
7--17\,{\micron} the uncertainty in the value of the crystallinity decreases to 5\,\%. 
We want to note, however, that probably the higher the crystallinity the easier it is to determine its
value regardless of the noise level. This is caused by the sharp peaks of the crystalline material 
which are easier to identify than the broad feature of the amorphous silicates. 

Our tests showed that the optimal wavelength range of the fit is 7--14\,{\micron} for
the ST and TT method assuming the mass absorption coefficients used in this paper. 
The width of this wavelength interval is limited by the similarities
in the optical data, which requires the broadest possible interval and the assumption
of a single average temperature in the feature and in the continuum, which acts in the 
opposite direction. The optimal wavelength range for the TLTD method has been
found to be 7--17\,{\micron}. The application of this broader wavelength range, compared
to the simpler methods, resulted in a better estimation of the dust composition. 
The 7--17\,{\micron} interval is broad enough to increase the differences among the 
mass absorption coefficients. On the other
hand, the temperature distribution is a considerably better estimation of the temperature
structure compared to an average temperature. Although, the 
TLTD method provided a better estimation of the dust composition than the other
methods, it has been found to be too simple to fit even broader wavelength intervals
(e.g.,\ the full {\it Spitzer} IRS domain, 5--35\,{\micron}). One should therefore
use the TLTD method for a first estimation of the dust composition by fitting the 
10\,{\micron} wavelength range, since a more realistic description of the radial
temperature structure and changes in the dust composition is necessary to fit an even 
broader wavelength range. Another possibility would be to apply the TLTD method
for the outer disk (e.g., 20-40\,{\micron} interval) and the longer wavelength part of the
Spitzer spectra and fit this range separately. 

On the basis of the results of our study, we propose the following recipe for the analysis
the silicate emission features.
\begin{enumerate}
\item {\it Set of mass absorption coefficients.} First, one should investigate how degenerate the 
  fitting problem is, i.e., how the mass absorption coefficients of a given dust species can
  be reproduced by the linear combination of that of the other dust species. 
  If there are two dust species with very similar mass absorption coefficients, then one should 
  handle them together, since they cannot be distinguished in the fitting procedure\footnote{In a real
  radiative transfer calculation this may be possible since the optical properties of different 
  materials are never completely identical over a large wavelength range.}. 

\item {\it Global shape of the SED and the disk structure.} The results of our study show that the
  applied assumption for the temperature structure of the disk is of fundamental importance.
  Therefore, one should investigate the global shape of the SED of the analyzed source in order
  to make a reasonable assumption for the disk structure. If the global shape of the SED can be 
  described with a disk model of \citet{ref:cg97}, \citet{ref:menshchikov_1997} or that of \citet{ref:ddn2001}, then one should 
  use the TLTD method for the analysis of the silicate feature. The TLTD method also has its 
  limitation. In its presented implementation it can handle only disks with a continuous surface density distribution. 
  For instance, if the disk has a large gap in the region where the silicate emission comes from, 
  then one has to modify the current form of the TLTD method by excluding this temperature range
  from the fit. 
  
\item {\it Spectral decomposition with the TLTD method.} Before the TLTD method is applied one 
  has to choose the wavelength range for the fit. We propose a wavelength range of 7--17\,{\micron} 
  for space-based observations (with the presented dust model), while for ground based observations 
  the wavelength domain is
  limited by the Earth's atmosphere. The continuum should not be subtracted from the spectrum,
  since this can introduce an additional uncertainty in the results, as our results have shown
  in the case of the COS method. We suggest to use the TLTD method in the form of integration
  over temperatures instead of radii in order to avoid assumptions for the outer disk radius.
  
\item {\it Interpretation.} The uncertainties in the resulting dust parameters can
  be estimated by a Monte Carlo type of error estimation from the observational uncertainties.
  For spectra with low signal-to-noise ratio (few tens) the fitted mass fraction of a single dust
  component can be very uncertain. Therefore we propose to use robust statistics, like the mass-averaged
  grain size or the crystallinity, for low S/N data in order to characterize the dust properties.
  Our test shows that for a S/N of 100 one can expect an uncertainty in the value of the crystallinity 
  of 11\,\% in terms of the total silicate mass for ground based observations (8--13\,$\mu$m). The
  expected uncertainty for space based observations (7--17\,$\mu$m) is 5\,\% for the same S/N level.
  
  During the interpretation, 
  one also should take into account the systematic errors in the fitting method. Using
  the optimal wavelength range for the fit, the TLTD method overestimates the mass-averaged 
  grain size of the amorphous silicates by 50$\pm$30\,\%, while the total crystallinity
  is overestimated by 8$\pm$7\,\%. The increase in the noise level, as well as a narrower
  wavelength range (8--13\,{\micron}) results in larger mass-average grain sizes for all dust 
  species and higher crystallinity values than the real values.

\end{enumerate}

\acknowledgments

Th. Henning was supported in part by the National Science Foundation under grant PHY05-51164 (at KITP).
We thank an anonymous referee for careful reading the paper and excellent suggestions for improving the text.
\appendix

\section{Disk models}
\label{sec:app_disk_models}

For the moderately flared and flat disks we use the following density distribution
\begin{equation}
\rho(r,z) = \frac{\Sigma(r)}{H_p(r)\sqrt{2\pi}}exp\left(-\frac{z^2}{{2H_p(r)}^2}\right), 
\end{equation}
where
\begin{equation}
\Sigma(r) = \Sigma_0\left(\frac{r}{R_{in}}\right)^{-p}
\end{equation}
and
\begin{equation}
\frac{H_p(r)}{r} = \frac{H_p(R_{in})}{R_{in}}\cdot\left(\frac{r}{R_{in}}\right)^{\alpha}.
\label{eq:hr}
\end{equation}
Here $H_P$ is the pressure scale height in the disk and $\alpha$ is the flaring index. 
For the puffed-up inner rim we applied a similar model than that of \citet{ref:ddn2001}.
However, the scale height of the inner rim was an adjustable parameter. It has been set
in such a way, that the shadowed region of the disks relative to their inner radius have
to be the same for all models in a given series. 

\bibliographystyle{aa}
\bibstyle{aa}

\bibliography{ms}

\clearpage

\begin{figure*}
\includegraphics[angle=90, width=8.2cm]{f1.eps}
\caption{Mass absorption coefficients of the dust species used in the tests. All dust 
species were used with three sizes: 0.1\,{\micron} (solid line), 1.5\,{\micron} (dotted line), 
6.0\,{\micron} (dashed line)}
\label{fig:opacities}
\end{figure*}

\clearpage

\clearpage

\begin{deluxetable}{lccc}
\tablecaption{Main parameters of the 2D RT disk models.}
\tablehead{\colhead{} & \colhead{Herbig Ae star} & \colhead{T Taur star} & \colhead{Brown Dwarf}}
\startdata
$L_{\star}$ [$L_{\odot}$] & 30   & 1.4  & 0.035 \\
$T_{\star}$ [K]          & 9500 & 4000 & 2500 \\
$M_{\star}$ [$M_{\odot}$] & 3   & 0.55  & 0.08 \\
$R_{in}$ [AU]            & 0.4   & 0.087  & 0.014 \\
$R_{out}$ [AU]           & 400   & 180  & 87 \\
\enddata
\label{tab:mod_prop}
\end{deluxetable}

\begin{deluxetable}{llllll}
\tablecaption{Overview of dust species used. For each component we specify
 its lattice structure, chemical composition, shape and reference to the 
 laboratory measurements of the optical constants. For the homogeneous
 spheres we used Mie theory to calculate the opacities. For the inhomogeneous
 spheres, we used the distribution of hollow spheres \citep[DHS;][]{ref:min_2005}, to simulate
 grain deviating from perfect symmetry.}
\tabletypesize{\small}
\tablewidth{0pt}
\tablecolumns{6}
\tablehead{\colhead{\#}&\colhead{Species}&\colhead{State}&\colhead{Chemical}&\colhead{Shape}&\colhead{Ref} \\
 \colhead{}& \colhead{}&\colhead{} & \colhead{ formula}  &  \colhead{}  & \colhead{} } 

\startdata
1 & Amorphous silicate  & A  &  MgFeSiO$_{4}$       & Homogeneous sphere   & (1)  \\
  & (Olivine stoichiometry)&  &                     &                &  \\
2 & Amorphous silicate  & A  &  MgFeSi$_{2}$O$_{6}$ & Homogeneous sphere   & (1)  \\
 & (Pyroxene stoichiometry)&  &                     &                &  \\
3 & Forsterite   & C  &  Mg$_{2}$SiO$_{4}$   & Hollow sphere & (2)   \\
4 & Clino Enstatite    & C  &  MgSiO$_{3}$         & Hollow sphere & (3) \\
5 & Silica       & A  &  SiO$_{2}$           & Hollow sphere & (4)  \\
6 & Amorphous Carbon & A & aC & Homogeneous sphere & (5)  \\
\tableline
\enddata
\label{tab:species}
\tablerefs{(1) \citet{dorschner1995}; (2) \citet{servoin1973}; (3) \citet{jaeger1998}; (4) \citet{Henning1997}; (5) \citet{preibisch1993}}
\end{deluxetable}

\begin{deluxetable}{lcccccc}
\tablecaption{Result of the spectral decomposition of the spectrum of a Herbig Ae disk
with an inclination of 45$^{\circ}$. Fitted wavelength domain : 7--14\,{\micron}.
Similar tables for all fits can be found in the electronic edition of the Journal.}
\tabletypesize{\scriptsize}
\tablehead{
\colhead{Name} & \colhead{Size} 
& \colhead{M.Fraction} & \colhead{M.fraction} & \colhead{M.Fraction} 
& \colhead{M.fraction} & \colhead{M.fraction} \\
     &  \colhead{[$\mu$m]}    & \colhead{Real [\%]} & \colhead{COS [\%]}
     &  \colhead{ST [\%]}     &  \colhead{TT [\%]}  & \colhead{TLTD [\%]}
}
\startdata
Olivine (A)     &   0.1   & 15.0 &  0.0 & 9.4 & 14.4  & 13.5  \\
Olivine (A)     &   1.5   & 7.0  & 19.5 & 0.0 &  0.0  &  6.8  \\
Olivine (A)     &   6.0   & 2.0  &  0.0 & 0.0 &  0.0  &  4.3  \\
Pyroxene (A)    &   0.1   & 15.0 &  0.0 & 1.5 & 44.6  & 14.4  \\
Pyroxene (A)    &   1.5   & 7.0  & 14.8 & 0.0 &  0.0  &  7.4  \\
Pyroxene (A)    &   6.0   & 2.0  &  0.0 & 0.0 &  0.0  &  2.8  \\
Forsterite (C)  &   0.1   & 2.0  &  0.1 & 1.7 &  2.5  &  1.9  \\
Forsterite (C)  &   1.5   & 1.0  &  0.0 & 0.0 &  1.4  &  1.0  \\
Forsterite (C)  &   6.0   & 0.0  & 28.5 & 0.0 &  0.0  &  0.1  \\
Enstatite  (C)  &   0.1   & 2.0  &  0.0 & 1.6 &  2.6  &  1.9  \\
Enstatite  (C)  &   1.5   & 1.0  &  1.7 & 2.2 &  0.1  &  0.9  \\
Enstatite  (C)  &   6.0   & 0.0  & 11.2 & 3.8 &  7.6  &  0.9  \\
Silica (A)      &   0.1   & 0.0  &  0.0 & 0.0 &  0.0  &  0.0  \\
Silica (A)      &   1.5   & 0.0  &  0.0 & 0.0 &  0.0  &  0.0  \\
Silica (A)      &   6.0   & 0.0  & 10.0 & 0.0 &  0.0  &  0.0  \\
Carbon (A)      &   0.1   & 46.0 & 14.0 & 1.6 &  0.0  & 43.8  \\
Carbon (A)      &   1.5   & 0.0  &  0.0 & 0.0 &  0.0  &  0.0  \\
Carbon (A)      &   6.0   & 0.0  &  0.0 &78.2 & 26.7  &  0.0  \\
\tableline
Average grain size              &  &  1.00 & 1.5 & 0.10 &  0.10 &  1.35 \\
Total crystallinity             &  &  6.00 & 41.5 & 9.30 & 14.20 &  6.79 \\

T$_{\rm f}$ [K]                 &  &    -    &- & 380.0  &  190.0  &    -    \\
T$_{\rm c}$ [K]                 &  &    -    &- & 380.0  &  615.0  &    -    \\
T$_{\rm rim}$(R$_{\rm in}$) [K]  &  &    -    &- &   -    &    -    & 1408.0  \\
T$_{\rm atm}$(1\,AU) [K]        &  &    -    &- &   -    &    -    &  755.6  \\
T$_{\rm mid}$(1\,AU) [K]        &  &    -    &- &   -    &    -    &  171.9  \\
p$_{\rm rim}$                   &  &    -    &- &   -    &    -    & -1.39   \\
p$_{\rm atm}$                   &  &    -    &- &   -    &    -    & -0.35   \\
p$_{\rm mid}$                   &  &    -    &- &   -    &    -    & -0.10   \\
Reduced $\chi^2$                &  &    -    &301.6 &2697.0 & 33.19 & 0.051 \\
\enddata
\label{tab:results}
\end{deluxetable}

\end{document}